\documentclass[aoas]{imsart}

\RequirePackage{amsthm,amsmath,amsfonts,amssymb}
\RequirePackage[authoryear]{natbib}
\RequirePackage[colorlinks,citecolor=blue,urlcolor=blue]{hyperref}
\RequirePackage{graphicx}

\startlocaldefs
\theoremstyle{plain}

\newtheorem{theorem}{Theorem}[section]

\theoremstyle{remark}
\newtheorem{definition}[theorem]{Definition}

\endlocaldefs

\usepackage{xcolor}
\usepackage[pagewise]{lineno}
\usepackage{rotating,multicol,multirow,array,styles}
\usepackage{float,tabularx,longtable}
\usepackage[ruled,vlined,linesnumbered]{algorithm2e}

\newcommand\blfootnote[1]{%
  \begingroup
  \renewcommand\thefootnote{}\footnote{#1}%
  \addtocounter{footnote}{-1}%
  \endgroup
}

\newcommand{\change}[1]{{\leavevmode\color{black}{#1}}}

\newcommand{\vvec}[1]{\mathrm{vec} \left( #1 \right)}
\newcolumntype{Y}{>{\centering\arraybackslash}X}
\usepackage{mathtools}

\makeatletter
\renewcommand\hyper@natlinkbreak[2]{#1}
\makeatother
\hypersetup{
	pdfborder = {0 0 0}
}

\usepackage{enumerate}
\usepackage[title,toc,titletoc,page]{appendix}
\renewcommand{\appendixname}{Supplement}  

\begin{document}

\begin{frontmatter}
\title{Tree-Regularized Bayesian Latent Class Analysis for Improving Weakly Separated Dietary Pattern Subtyping in Small-Sized Subpopulations}
\runtitle{Tree-Regularized Bayesian Latent Class Analysis}

\begin{aug}
\author[A]{\fnms{Mengbing}~\snm{Li}\ead[label=e1]{mengbing@umich.edu}\orcid{0000-0002-2264-8006}},
\author[B]{\fnms{Briana}~\snm{Stephenson\textsuperscript{$\ast$}}\ead[label=e2]{bstephenson@hsph.harvard.edu}\orcid{0000-0002-6147-1039}}
\and
\author[A]{\fnms{Zhenke}~\snm{Wu\textsuperscript{$\ast$}}\ead[label=e3]{zhenkewu@umich.edu}\orcid{0000-0001-7582-669X}}
\address[A]{Department of Biostatistics, University of Michigan\printead[presep={,\ }]{e1,e3}}
\address[B]{Department of Biostatistics, Harvard University\printead[presep={,\ }]{e2}}
\end{aug}

\begin{abstract}
Dietary patterns synthesize multiple related diet components, which can be used by nutrition researchers to examine diet-disease relationships. Latent class models (LCMs) have been used to derive dietary patterns from dietary intake assessment, where each class profile represents the probabilities of exposure to a set of diet components. However, LCM-derived dietary patterns can exhibit strong similarities, or weak separation, resulting in numerical and inferential instabilities that challenge scientific interpretation. This issue is exacerbated in small-sized subpopulations. To address these issues, we provide a simple solution that empowers LCMs to improve dietary pattern estimation. We develop a tree-regularized Bayesian LCM that shares statistical strength between dietary patterns to make better estimates using limited data. This is achieved via a Dirichlet diffusion tree process that specifies a prior distribution for the unknown tree over classes. Dietary patterns that share proximity to one another in the tree are shrunk towards ancestral dietary patterns \textit{a priori}, with the degree of shrinkage varying across pre-specified food groups. Using dietary intake data from the Hispanic Community Health Study/Study of Latinos, we apply the proposed approach to a sample of 496 US adults of South American ethnic background to identify and compare dietary patterns. 
\end{abstract}


\end{frontmatter}

\blfootnote{\textsuperscript{$\ast$}Co-senior authors. }

\section{Introduction} \label{sec:introduction}

\subsection{Motivation} \label{sec:inrtoduction:motivation}
Dietary patterns (DPs) refer to the quantities, proportions, variety, or combination of different foods, drinks, and nutrients, and the frequency with which they are consumed \citep{dietary2020ScientificReport}.
\change{Understanding the heterogeneity of dietary consumption behaviors provides vital insights into the relationship between diet and health outcomes \citep{mokdad2018state}, enabling tailored nutrition interventions for subgroups with distinct DPs.}

\change{
In the Hispanic Community Health Study/Study of Latinos \citep[HCHS/SOL; ][]{lavange2010SampleDesign}, 24-hour dietary recalls were collected at baseline to participants to evaluates foods consumed within the past 24-hours. Due to the open-ended format, thousands of foods are observed and are often aggregated into categorical daily serving levels of a smaller food item set for analysis \citep{stephenson2023identifying,uzhova2018generic,sotres2010latent}. }
While DP can be analyzed on individual foods or nutrients, nutritionists argue for a holistic approach because foods are not consumed in isolation and nutrients have synergistic effects \citep{kant2004DietaryPatterns}. 
Given the large number of foods to jointly analyze, dimension reduction techniques are often used to understand dietary behaviors.

\change{
Existing studies often identify global DPs across an entire cohort, with limited attention to localized variation in smaller demographic subpopulations defined by study site or ethnicity. For example, \citet{mattei2016diet} found that South American (SA) participants in HCHS/SOL deviated from many DPs shared by other ethnic groups. However, due to its relatively smaller sample size, the SA subpopulation is statistically overwhelmed, making it difficult for the global model to recognize localized DPs. While methods exist to capture local deviations \citep{vito2022SharedEthnic,stephenson2020empirically}, they often exclude small site-ethnic subpopulations (e.g., Bronx participants of Central American background, $N=217$). Alternative analyses pool across study sites into subpopulations defined by ethnicity \citep[e.g.,][]{maldonado2021dietary}. 
However, excluding or pooling across small subpopulations is ad hoc and less ideal because we may ignore demographic nuances that drive dietary behavior differences. Developing stable and accurate DP estimation methods for small-sized subpopulations is crucial for both statistical rigor and health equity.

DP estimation in small subpopulations faces unique statistical challenges. First, DPs may only be distinguished by a subset of food items, particularly in culturally or regionally homogeneous groups \citep{mattei2016diet,stephenson2020empirically}. Capturing these nuanced differences that may be masked by dominant patterns is critical, as even minor variations in DPs can significantly impact health outcomes, such as cardiovascular risks and obesity \citep{farmer2020cooking}. 
Merging similar DPs obscures these implications and limits tailored nutritional interventions, not to mention the difficulty of specifying a meaningful similarity threshold for merging. 
While distinguishing these nuances may increase computational cost, the potential health insights justify the effort.
Second, DPs exhibit varying degrees of distinction across major food groups. For instance, fruits and vegetables often reveal clearer differences between DPs compared to sweets and fats \citep{sotres2010latent}. Recognizing such varying differences can improve estimation accuracy and highlight meaningful dietary behaviors.
Finally, small sample sizes exacerbate these two challenges. The meaning of a small or large sample size is relative to how distinct the DPs are in a dataset.
The significance of DP variability depends on sample size—smaller variations may be more meaningful in a small sample compared to a large one. An ideal method would guard against steep deterioration of inferential quality and numerical stability in smaller samples, as standard methods tend to fail in these situations.
}

\subsection{Existing Methods} \label{sec:intro:existingmethod}

\change{
Latent class models \citep[LCMs;][]{lazarsfeld1950logical} are widely used for deriving population-based DPs from multivariate categorical food exposure data, as in our 24-hour dietary recalls \citep{sotres2010latent,huh2011identifying,leech2017temporal,garcia2019dietary,park2020latent}. 
Other approaches for DP analysis based on continuous intake measurements like nutrient content include factor analysis \citep{schulze2003approach,mattei2016diet}, principal component analysis \citep{hearty2008comparison}, and K-means clustering \citep{stricker2013dietary}. 
LCMs partition the population into $K$ mutually exclusive latent classes, summarizing DPs as the class-specific conditional probabilities of different exposure levels to food items. An individual assigned to the $k$-th class has a probability vector, $\bm \theta_{j,k} = (\theta_{j,k,1}, \ldots, \theta_{j,k,d_j})^\top$ such that $0 < \theta_{j,k,h} < 1$ and $\sum_{h=1}^{d_j} \theta_{j,k,h} = 1$, that represents the probabilities of $d_j$ exposure levels to item $j$.
}

\change{
LCMs can be estimated via maximum likelihood or Bayesian methods, but classical approaches struggle with detecting nuanced differences between weakly separated patterns, where inter-pattern distances (e.g., Mahalanobis distance, intraclass correlation coefficient, $R$-square entropy) are small \citep{lubke2007PerformanceFactor,rights2016relationship,ramaswamy1993empirical}. 
Under weak separation, challenges such as convergence failure, poor model fit, and poor empirical identifiability arise, especially in small samples \citep{park2018RecommendationsSample,weller2020LatentClass}. 
Consider a toy example of $K = 3$ classes with equal class proportions, based on $J$ items with binary exposure levels.
The three patterns $\{\btheta_{j,1} = (0.8, 0.2), \btheta_{j,2} = (0.2, 0.8), \btheta_{j,3} = (0.5, 0.5); j = 1, \ldots, J\}$ are strongly separated, compared to $\{\btheta_{j,1} = (0.65, 0.35), \btheta_{j,2} = (0.35, 0.65), \btheta_{j,3} = (0.5, 0.5); j = 1, \ldots, J\}$ which are weakly separated.
During estimation of the latter case, the classical LCM would consider only two classes exist and the remaining class has approximately zero proportions, and therefore fail to estimate the three patterns correctly.
In fact, this phenomenon is referred to ``the curse of dimensionality'' in Bayesian mixture models \citep{chandra2023escaping}, where too many or too few clusters occur when the number of items $J$ increases. 
This happened to our DP analysis using the classical LCM to the South American subpopulation from the Hispanic Community Health Study/Study of Latinos (HCHS/SOL), as discussed in Section \ref{sec:data:results}.
Additionally, classical LCMs do not account for varying degrees of separation among major food groups.

Prior studies have used flexible models to improve DP analysis. For example, robust profile clustering \citep[RPC,][]{stephenson2020RobustClusteringa} allows for subpopulation-specific local patterns may deviate from global patterns overarching the entire population. However, RPC is not designed for weakly separated DPs and is constrained by sample size to effectively estimate local patterns. Subpopulations ranging from 300 to 3000 individuals yielded only one or two local patterns, likely due to distinct but weakly separated patterns being inadvertently merged \citep{stephenson2020empirically}. 
Other studies have incorporated covariates to improve LCM's ability to differentiate between subjects and enhance latent class interpretability, but there are several caveats. 
When weakly separated classes and small sample sizes are present, class profile and covariate effect estimation are often compromised \citep{li2017investigating,asparouhov2014auxiliary}. Covariate selection also becomes challenging as the number of potential covariates grows. Moreover, in this paper we study DPs as an exposure analysis, as opposed to assessing the covariate effects on DPs as an outcome analysis. Our exposure analysis aims to identify dietary behaviors shared by latent subgroups at the population level, and hence direct covariate inclusion is less central to our primary goals.
}

\subsection{Main Contributions}	\label{sec:intro:contributions}
\change{
We introduce a tree-regularized LCM, a general framework that improves parameter estimation by sharing statistical information across classes. The proposed model addresses weak class separation in small sample sizes by (1) incorporating an unknown tree to guide shrinkage, and (2) accounting for varying degrees of shrinkage across major food groups. The Dirichlet diffusion tree (DDT) process \citep{neal2003density} is used to specify a prior on the class profiles on the leaves of an unknown tree (hence termed ``DDT-LCM''), encouraging DPs closer on the tree to exhibit more similarities. 
The degrees of separation by major food groups are modeled by group-specific diffusion variances.
}

\change{
While various strategies induce parameter dependency (e.g., shrinkage priors in regression models), we employ a tree-structured prior for its power in organizing hierarchical relationships between entities and providing interpretable insights. 
In DP analysis, trees are intuitive for comparing and visualizing patterns across a large number of food items. Trees have also been applied in other scientific contexts. For example, a known tree structure can be incorporated into models to boost statistical powers, as seen in phylogenetic trees for studying zoonotic disease origins \citep{li2023IntegratingSample}, hierarchical hospital discharge codes for detecting associations between air pollution and cardiovascular outcomes \citep{thomas2020estimating}, and hierarchical topic models \citep{blei2003HierarchicalTopica,roy2006LearningAnnotated,zavitsanos2011non,weninger2012document}.
Even if no explicit hierarchy is present, component distributions in mixture models may share parameters. 
Certain LCM-derived DPs can share consumption similarities across many food items.
However, classical mixture models impose independent priors on components, ignoring structures and requiring more data for reliable estimation \citep{neal2003density,knowles2015pitman}.
One solution is to impose a sparse prior that identifies relevant items to a pattern and sets irrelevant items at some baseline distribution shared across patterns \citep{hoff2005subset,zhou2015bayesian}. Yet the fixed baseline distribution for DPs requires extensive nutrition knowledge and is difficult to specify.
}

\change{
Our approach differs significantly from existing tree-based methods. Latent class tree \citep[LCT,][]{van2019latent} hierarchically splits classes in a stepwise manner for interpretability but assumes equal branch lengths and focuses only on tree topology.
In contrast, DDT-LCM aims to improve empirical identifiability of weakly separated classes, allowing varying branch lengths for different degrees of shrinkage to reflect class similarities. 
Latent tree analysis \citep{zhang2004hierarchical}, on the other hand, builds a tree to explain between-item correlations, serving a fundamentally different goal than our framework.
}


The rest of the paper is organized as follows. Section \ref{sec:model}
introduces the standard LCM and the proposd DDT-LCM framework.
Section \ref{sec:inference} derives the posterior sampling algorithm, and Section \ref{sec:simulation} evaluates model performance via simulation studies. Section \ref{sec:data} applies DDT-LCM, via an R package \href{https://cran.r-project.org/web/packages/ddtlcm/index.html}{\texttt{ddtlcm}}, to identify food consumption patterns obtained from 24hr dietary recalls, of South American adults. Section \ref{sec:discussion} discusses study limitations and future directions. 

\section{Model Formulation} \label{sec:model}
Throughout this paper, ``major food group'' refers to the pre-specified nutritional groups of food items, and ``class'' refers to individuals' unknown dietary pattern memberships. For a positive integer $K$, we denote the set of positive integers up to $K$ by $[K] = \{1, \ldots, K\}$. Also denote the $K$-dimensional probability simplex by $\Delta^{K-1} = \{(\pi_1, \ldots, \pi_K): \pi_k \geq 0, \sum_{k=1}^{k=k} \pi_k = 1\}$.

\subsection{Latent Class Models} \label{sec:model:classicalLCM}

\change{
The dietary assessment dataset $\bY = (Y_{i,j})_{N \times J}$ queries $N$ individuals on exposure to $J$ food items. For individual $i \in [N]$, $Y_{i,j} \in [d_j]$ takes one of the $d_j$ categories and represents the observed exposure level to the $j$-th item.
We also have a pre-specified grouping vector $\br = (r_1, \ldots, r_J) \in [G]^J$, where $r_j = g$ if item $j$ (e.g., meat) is categorized into the $g$-th major group (e.g., lamb, poultry). 
The number of items in group $g$ is $J_g$ such that $J = \sum_{g=1}^G J_g$. 
A $K$-class LCM assumes that the $N$ individuals can be partitioned into $K$ latent classes, and that each class has distinct exposure behaviors to the $J$ items. 
The generative process of this model can be written as follows:
\begin{align} 
    Z_i \mid \bpi &\sim \mathrm{Categorical}_K (\bpi), i \in [N] \label{eq:classicalLCM:assignment} \\
    Y_{i,j} \mid Z_i = k, \btheta_{j,k} &\sim \mathrm{Categorical}_{d_j} \left( \btheta_{j,k} \right), k \in [K], j \in [J], \label{eq:classicalLCM:observation}
\end{align}
where $\bm \pi \in \Delta^{K-1}$ is the class prevalence vector, and $\btheta_{j,k} \in \Delta^{d_j-1}$ contains the conditional probability distribution of exposure levels to item $j$ in class $k$. 
As shown in \eqref{eq:classicalLCM:observation}, the observations $Y_{i,1}, \ldots, Y_{i,J}$ are conditionally independent given individual class assignment $Z_i$.
The $k$-th dietary pattern is characterized by the vector $\btheta_{k} = (\btheta_{1,k}, \ldots, \btheta_{J,k})^\top \in [0,1]^D$, where $D = \sum_{j=1}^{J} d_j$ is the total number of categories among all items. 
The dietary pattern matrix $$\bm \bTheta = 
\begin{pmatrix}
    \bm \theta_{1,1} & \cdots & \bm \theta_{1,K} \\
    \vdots & \ddots & \vdots \\
    \bm \theta_{1,K} & \cdots & \bm \theta_{J,K} \\
\end{pmatrix}_{D \times K}$$
and individual class assignment vector $\bZ = (Z_1, \ldots, Z_N)^\top$ are of our primary interest.
After specifying a prior $p(\bTheta)$ on the patterns, the joint distribution of $(\bZ, \bTheta)$ is
\begin{align}
p(\bZ, \bTheta \mid \bY) \propto \prod_{k=1}^{K} \left[ \pi_k \prod_{j=1}^{J} \prod_{c=1}^{d_j} \theta_{j,k,c}^{I \{ Y_{i,j} = c\}} \right]^{I \{ Z_i = k\} } p(\bTheta). \label{eq:lcm:llk}
\end{align}
Many choices for the prior are available, such as Dirichlet process \citep{ferguson1973bayesian,dunson2009nonparametric} and Pitman-Yor process \citep{pitman1997two}. Another common prior is Dirichlet-Multinomial, assuming independent prior Dirichlet distribution on the patterns across classes and items with 
\begin{equation} \label{eq:lcm:dirichlet:prior}
    \btheta_{j,k} \sim \mathrm{Dirichlet}_{d_j} \left( \balpha_{j,k} \right).
\end{equation}
The uniform corresponds to $\balpha_{j,k} = (1, \ldots, 1)$, and the noninformative Jeffery's prior corresponds to $\balpha_{j,k} = (1/2, \ldots, 1/2)$.


}

\begin{figure}[t]
    \centering
    \includegraphics[width=0.9\linewidth]{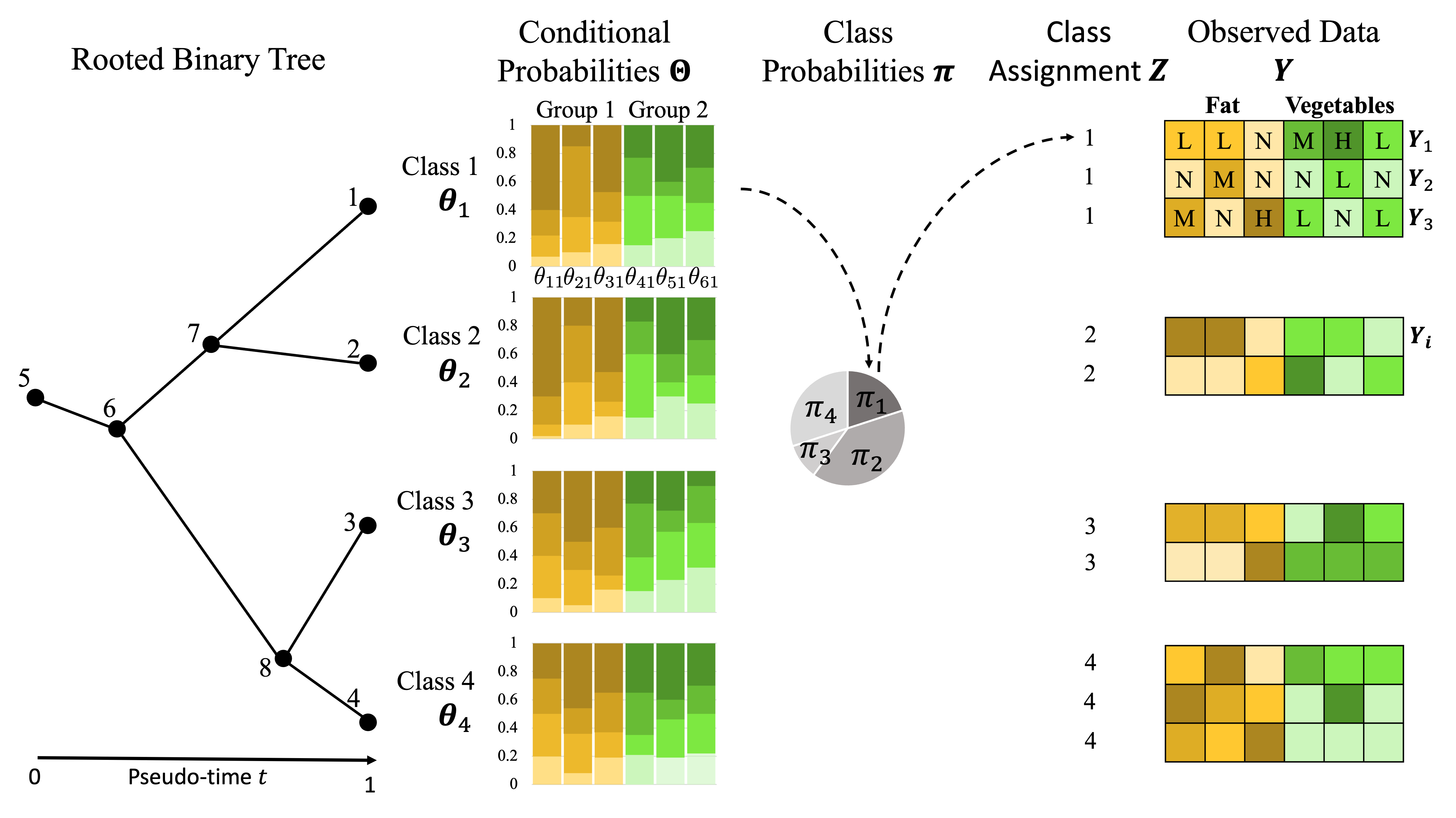}
    \vspace{-1em}
    \caption{Illustration of DDT-LCM with $K = 4$ classes, and $J = 6$ items categorized into $G = 2$ major food groups (fat and vegetables) in distinct colors. \textit{Right}: In the observed data matrix $\bY$ for $N = 10$ individuals, each cell records none (N), low (L), medium (M), or high (H) exposure level to a food item.
    \textit{Middle}: The dietary patterns are characterized by the conditional probabilities $\bTheta = (\theta_{j,c,k})$. Each element $\theta_{j,c,k}$ is represented by the bar height.
    \textit{Left}: A tree drawn from a DDT process to capture the hierarchical relationship over the dietary patterns. In our data analysis, this tree is unknown and needs to be estimated.}
    \label{fig:model_structure}
\end{figure}

\subsection{Proposed Prior on Dietary Patterns} \label{sec:model:proposed}

\change{
To improve empirical identifiability of the LCM under weakly separated classes, we enable the sharing of statistical strength between classes through specifying a tree-structured shrinkage prior on $\bTheta$. Specifically, we use the Dirichlet diﬀusion tree (DDT) process \citep{neal2003density,knowles2015pitman} as the prior;
hence we term our model ``DDT-LCM". In the following, we begin with an overview of our model and then discuss more technical details about parameters in the prior.

\subsubsection{Overview of DDT-LCM} \label{sec:model:prior}
\sloppy For item $j$ in class $k$, we define $\bm \eta_{j,k} = \left( \log \frac{ \theta_{j,k,1} }{ \theta_{j,k,d_j}}, \ldots, \log \frac{ \theta_{j,k,d_j-1} }{ \theta_{j,k,d_j}} \right) \in \RR^{d_j - 1}$ as the vector of log odds ratios of conditional exposure probabilities compared to the last category $d_j$. 
Let the logit-transformed patterns be the column concatenation of the $K$ patterns 
\begin{align} 
    \bm \eta = \begin{pmatrix}
        \bm \eta_{1,1} & \cdots & \bm \eta_{1,K} \\
        \vdots & \ddots & \vdots \\
        \bm \eta_{J,1} & \cdots & \bm \eta_{J,K} \\
    \end{pmatrix}_{D_0 \times K}, \text{ where } D_0 = D-J. \label{eq:notation:eta}
\end{align}
In particular, if all items are binary, then $D_0 = 2J - J = J$. 
We assume the following prior on the column vectors of $\bm \eta$:
\begin{align} \label{eq:ddt:prior}
\left( \bm \eta_1, \ldots, \bm \eta_K \right) \mid c, \bsigma^2 \sim \mathrm{DDT} (a(t; c), \bsigma^2),
\end{align}
where $a(t; c)$ is a divergence function on pseudo-time $t \in [0, 1]$ with parameter $c$, and $\bsigma^2 = (\sigma_1^2, \ldots, \sigma_G^2)^\top \in \RR_+^G$ is a vector of group-specific diffusion variance parameters. See \ref{sec:model:property} for details about $a(t; c)$ and $\bsigma^2$.
With this prior, the $K$ patterns $\{\bm \eta_k\}_{k=1}^K$ are viewed as drawn from a DDT process. 
The DDT induces a structured marginal covariance \textit{a priori} between the patterns, after integrating out all intermediate paths of the particles; specifically the column covariance matrix of $\bm \eta$, $\bSigma_{K \times K} = \mathrm{ColCov} \left( \bm \eta \right)$, embeds the tree structure induced by DDT.
Consequently, classes positioned closer on the tree exhibit less between-pattern  separation, reflected by larger off-diagonal correlation entries in $\bSigma$. A larger degree of separation between classes in major food groups $g$ is driven by a larger diffusion variance $\sigma_g^2$. 

Figure \ref{fig:model_structure} illustrates an example of DDT-LCM with $K = 4$ classes. Classes 3 and 4 share more similar dietary patterns and are consequently positioned closer on the tree. Moreover, the two classes are more distinguishable among items in the fat group than the vegetable group. 
Multivariate categorical responses on the right are realized through the LCM given the class probabilities and the tree-regularized conditional probabilities.

The DDT process flexibly specifies a joint prior on the unknown tree structure and the exposure probabilities. This prior provides regularization on the patterns and is particularly effective under weak class separation and small sample sizes, as we will demonstrate via extensive simulations in Section \ref{sec:simulation}. 


\begin{remark} \label{remark:dependency}
(Types of dependency)
The prior in \eqref{eq:ddt:prior} introduces marginal dependence among the column vectors of $\bm \eta$, leading to betwee-pattern dependence among columns of $\bTheta$. This prior essentially relaxes the prior independence in \eqref{eq:lcm:dirichlet:prior}. However, we still impose the conditional independence assumption in \eqref{eq:classicalLCM:observation}. This is reflected by the block-diagonal structure in the between-item row covariance of $\bTheta$, as will be shown in Proposition \ref{prop:marginal:theta} in Section \ref{sec:model:property}. Specifically, $\mathrm{RowCov} \left( \bm \Theta \right)$ of size $D_0 \times D_0$ contains $J$ diagonal blocks, where the $j$-th block contains the marginal prior covariance between the $d_j - 1$ non-baseline categories of item $j$. Methods for incorporating between-item dependence is out of the scope of this paper but have been considered in \cite{qu1996random,zhang2004hierarchical,hoff_extending_2007}. 
\end{remark}
}

\change{
\subsubsection{Description of DDT Process} \label{sec:model:ddt}
We next describe more details about the DDT as a prior in our LCM. Further background about the DDT process is included in Supplement \ref{supp:sec:ddt}. 
The DDT process provides a family of nonparametric distributions over $K$ exchangeable ``particles'' arising from a latent branching process on the pseudo-time interval $[0,1]$. The latent state of the $k$-th particle at time $t \in [0, 1]$ is denoted as $\bX_k(t) \in \RR^{D_0}$, where the dimension $D_0$ equals the number of rows in $\bm \eta$. 

From the generative perspective, $K$ particles are iteratively simulated via a self-reinforcement scheme; hence the name ``Dirichlet'' as in Dirhclet process \citep{ferguson1973bayesian}. The instantaneous probability of diverging from existing paths is parameterized by the divergence function $a(t; c), t \in [0, 1]$ as in \eqref{eq:ddt:prior}; see Supplement \ref{supp:sec:ddt} 
for the exact formula. The paths of these particles result in a $K$-leaf rooted binary tree with topology $\cT$ and branch times $\bB = \{t_k \in [0, 1]; k \in [2K] \}$, where the $K$ leaf nodes are labeled as $1, \ldots, K$, the root node as $K+1$, and the $K-1$ internal nodes as $K+2, \ldots, 2K$. In particular, the root node is at time $0$ such that $t_{K+1} = 0$.
In subsequent sections of the paper, we choose
\begin{align} \label{eq:divfunction}
a(t; c) = c / (1 - t), c > 0,
\end{align}
following \cite{neal2003density}, where $c$ is a ``smoothness" hyperparameter. Larger $c$ values place higher prior weights on earlier divergence, leading to shallower trees. 

Along the tree structure (topology and branch times), the latent states are generated from a multivariate Brownian motion scaled by variance $\bsigma^2$ as in \eqref{eq:ddt:prior}. 
Specifically, let $\bOmega$ be a $D_0 \times D_0$ diagonal matrix where the rows are indexed by the $D_0$ non-baseline categories across the $J$ items. For each $j \in [J]$ and $c \in [d_j-1]$, the $( \sum_{h=1}^{j-1} (d_h - 1) + c, r_j)$-th diagonal entry of $\bOmega$ equals $\sigma_j^2$.
The $k$-th particle will diffuse to $\bX_k(t + dt) = \bX_k(t) + \mathrm{MVN}(\bm 0_{D_0},  \bOmega \cdot dt)$ after infinitesimal time $dt$ once it reaches $\bX_k(t)$. 
Here $\mathrm{MVN}(\bmu, \bPsi)$ denotes the multivariate normal distribution with mean $\bmu$ and covariance $\bPsi$.  


Marginalized over intermediate paths between nodes, a draw from the distribution over a $K$-leaf DDT yields 
\begin{align}
\left\{ \cT, \left( t_k, \bm \eta_k \right)_{k=1}^{2K} \right\},
\end{align}
where $\bm \eta_k := \bX_k (t_k)$ are the latent states at nodes of the tree whose topology is given by $\cT$. For notation clarity, we emphasize that the matrix $\bm \eta$ in \eqref{eq:notation:eta} only contains latent states at leaf nodes $\bm \eta_1, \ldots, \bm \eta_K$. 
}

\change{
\subsubsection{Properties of DDT-LCM} \label{sec:model:property}
The role of the DDT as a prior can be viewed from two different perspectives. First, as stated in Section \ref{sec:model:prior}, the DDT prior imposes a marginal between-pattern covariance $\bSigma$ that embeds a tree structure. Second, the DDT can in fact be viewed as a tree-structured shrinkage prior if we consider conditional distributions of latent states in the generative process.

For the first perspective, the following proposition derives the marginal distribution of latent states $\left( \bm \eta_1, \ldots, \bm \eta_K \right)$ at leaf nodes. This marginal distribution is also essential for developing a sampling algorithm for posterior inference in Section \ref{sec:inference}. Let $\cM \cN_{D_0 \times K} (\bM, \bU, \bW)$ denote the matrix normal distribution of a $D_0 \times K$ random matrix with mean $\bM$, $D_0 \times D_0$ row covariance matrix $\bU$, and $K \times K$ column covariance matrix $\bW$.

\begin{proposition} \label{prop:marginalleaf}
Given a tree topology $\cT$ and branch times $\bB = \{t_k \in [0, 1]: k \in [2K] \}$ drawn from the DDT process with diffusion variances $\bsigma^2$ in \eqref{eq:ddt:prior}, the distribution of $\bm \eta$ marginalized over all intermediate stochastic paths between tree nodes is 
\begin{equation}
    \bm \eta \mid \cT, \bB, \bsigma^2 \sim \cM \cN_{D_0 \times K} \left( \bm 0, \bOmega, \bSigma \right), \label{eq:marginalleaf}
\end{equation}
where $\bSigma$ is a covariance matrix with entries $\bSigma_{k, l} = t_{\mathrm{MRCA} (k, l)}, k,l \in [K]$. Here MRCA$(k, l) \in \{K+2, \ldots, 2K\}$ is the most recent common ancestor of leaf nodes $k$ and $l$.
\end{proposition}

Proposition \ref{prop:marginalleaf} states that the DDT prior leads to a marginal matrix Gaussian distribution on the logit-transformed patterns centered at the the origin. The between-item row covariance matrix is diagonal with entries $\sigma_g^2$ for all non-baseline categories of items in group $g$. The between-class column covariance matrix $\bSigma$ encodes the tree structure with entries being the branch time of the MRCA of two leaf nodes. 
In this paper, we call $\bSigma$ a ``tree-structured covariance matrix''. Due to its one-to-one correspondence with a rooted binary tree, we may assess the distance between two DDT trees by the distance between the associated matrices (e.g., Frobenius norm of matrix differences; see Section \ref{sec:simulation}). Supplement \ref{supp:sec:treecov_example} 
provides more discussion about $\bSigma$, and an example to illustrate how the covariance structure depends on the tree topology and branch times.

\begin{remark} \label{remark:ceffects}
In the divergence function \eqref{eq:divfunction}, larger $c$ values lead to earlier divergence and shallower trees. Following Proposition \ref{prop:marginalleaf}, $\bSigma$ will have smaller off-diagonal entries and hence encourage smaller marginal between-pattern correlations \textit{a priori}.
\end{remark}
}

\change{
We next define a \textit{Matrix Logistic Normal} distribution that is useful for understanding the prior distribution on $\bTheta$.
\begin{definition} \label{def:matrix:logitnormal}
Suppose that $\bX = (x_{d,k})_{(D-1) \times K}$ is a random matrix following a matrix normal distribution $\cM \cN (\bM, \bU, \bW)$. We say that $\bV = (v_{d,k})_{D \times K}$ is a Matrix Logistic Normal distribution if $v_{d,k} = \frac{ \exp(x_{d,k}) }{ 1 + \sum_{c=1}^{D-1}\exp(x_{c,k})}$ for $d \in [D-1]$ and $v_{D,k} = \frac{ 1 }{ 1 + \sum_{c=1}^{D-1}\exp(x_{c,k})}$, $k \in [K]$. 
We denote the distribution of $\bV$ as $\mathrm{MLN}(\bM, \bU, \bW)$.
\end{definition}

Definition \ref{def:matrix:logitnormal} is a matrix generalization of the vector logistic normal distribution discussed in \cite{atchison1980logistic}. 
The density function and properties of the MLN distribution are provided in Section  \ref{sec:matrixlogitnormal} 
of the Supplement.
The following proposition gives the prior marginal distribution of $\bTheta$ in our model.

\begin{proposition} \label{prop:marginal:theta}
Given the distribution of $\bm \eta$ in \eqref{eq:marginalleaf}, the prior distribution of $\bTheta$ marginalized over all intermediate stochastic paths between tree nodes is
\begin{align}
    \bTheta \mid \cT, \bB, \bsigma^2 \sim \prod\limits_{j=1}^{J} \mathrm{MLN}_{d_j \times K} \left( \bm 0, \sigma_{r_j}^2 I_{d_j-1}, \bSigma \right). \label{eq:theta:distribution:logitnormal}
\end{align}
In other words, for each $j \in [J]$ the $d_j \times K$ matrix $\btheta_{j,\cdot}$ is MLN-distributed with diagonal row covariance $\sigma_{r_j}^2 I_{d_j-1}$ and tree-structured column covariance $\bSigma$. In addition, $\btheta_{1,\cdot}, \ldots, \btheta_{J,\cdot}$ are independent across all $j$'s.
\end{proposition}
}

\change{
The joint distribution of the tree structure and patterns marginalized over all intermediate stochastic paths between tree nodes, $P \left( \bTheta, \cT, \bt \mid c, \bsigma^2 \right)$, is given by equation \eqref{eq:ddt:jointdensity} in Section  \ref{supp:sec:marginalprior} 
of the Supplement. This distribution is used to derive an MCMC algorithm for posterior inference in Section \ref{sec:inference}.
}

\change{
For the second perspective, we consider two leaf nodes $k, l \in [K]$ satisfying $pa(k) \neq \mathrm{MRCA}(k, l)$ along a fixed tree structure $(\cT, \bB)$. For example, in the tree shown in Figure \ref{fig:model_structure}, we may consider $k = 1$ and $l = 3$, so that $pa(k) = 7, pa(l) = 8$, and $\mathrm{MRCA}(k, l) = 6$.
Due to properties of Brownian motion, the conditional distributions of $\bm \eta_k$ and $\bm \eta_l$ given the latent states of their parent nodes are
\begin{align} 
    \begin{split} \label{eq:model:conddist}
        \bm \eta_{k} \mid \bm \eta_{pa(k)}, t_{pa(k)}, t_k=1 &\sim \cN \left( \bm \eta_{pa(k)}, (1 - t_{pa(k)}) \bOmega \right) \\
        \bm \eta_{l} \mid \bm \eta_{pa(l)}, t_{pa(l)}, t_l=1 &\sim \cN \left( \bm \eta_{pa(l)}, (1 - t_{pa(l)}) \bOmega \right).    
    \end{split}
\end{align}
Similarly, consider the MRCA, denoted as node $h \in \{K+2, \ldots, 2K\}$, of nodes $pa(k)$ and $pa(l)$. The conditional distributions of $\eta_{pa(k)}$ and $\eta_{pa(l)}$ given the latent state of node $h$ are
\begin{align}
    \begin{split} \label{eq:model:conddist:hierarchy}
        \bm \eta_{pa(k)} \mid \bm \eta_{h}, t_h, t_{pa(k)} &\sim \cN \left( \bm \eta_{h}, (t_{pa(k)} - t_h) \bOmega \right) \\
        \bm \eta_{pa(l)} \mid \bm \eta_{h}, t_h, t_{pa(l)} &\sim \cN \left( \bm \eta_{h}, (t_{pa(l)} - t_h) \bOmega \right).
    \end{split}
\end{align}
Notice that \eqref{eq:model:conddist} and \eqref{eq:model:conddist:hierarchy} specify a hierarchical prior on $\bm \eta$ such that $\bm \eta_{k}$ and $\bm \eta_{l}$ are shrunk towards $\bm \eta_{h}$. Therefore, the DDT prior in \eqref{eq:ddt:prior} can also be viewed as a shrinkage prior.

Moreover, the proposed DDT-LCM has an intuitive interpretation of the tree over patterns $\bTheta$. The root node represents a ``root latent class'' encompassing all individuals, which then successively splits into child classes when branching occurs on the tree. 
As a result, each internal node can be viewed as an ``ancestral latent class'' with logit-transformed ``ancestral patterns'' centered at the parent latent state. This ensures similarity between the child patterns. After $K-1$ splits, the tree yields $K$ leaf classes, whose latent states enter the model likelihood in \eqref{eq:lcm:llk}.
}

\paragraph*{Priors for other parameters} 
We place conjugate priors on the remaining model parameters. The prior on the class probability vector $\bpi$ is a Dirichlet distribution $\bpi \sim \mathrm{Dirichlet} (\alpha_{\pi_1}, \ldots, \alpha_{\pi_K})$. For the hyperparameter $c$ in the divergence function $a(t)$, we choose a gamma prior with shape $\alpha_c$ and rate $\beta_c$ respectively, i.e. $c \sim G(\alpha_c, \beta_c)$. For the diffusion variances, we place inverse-gamma priors $\sigma^2_g \sim IG \left( \alpha_{\sigma^2_g}, \beta_{\sigma^2_g} \right)$ for $g \in [G]$.

\section{Algorithm for Posterior Inference} \label{sec:inference}
The full posterior distribution for all the unknowns $\bPhi=\{\cT, \bt, \bm \eta, c, \bsigma^2, \bZ, \bpi\}$ is
\begin{align} \label{eq:posterior:factors}
    & p\left(\bPhi \mid \bY \right) \propto \ p(\cT, \bt \mid c) p(\bm \eta \mid \cT, \bt, \bsigma^2) p(\bZ \mid \bpi) p(\bY \mid \bZ, \bm \eta ) p(c) p(\bsigma^2) p(\bpi).
\end{align}
Note we have marginalized over the internal node latent states to focus on leaf latent states $\bm \eta$ that directly parameterize the class profiles.
We consider a hybrid Metropolis-Hastings-within-Gibbs algorithm (Algorithm \ref{alg:inference}) to sample from (\ref{eq:posterior:factors}) in three steps: 
(a) a Metropolis-Hastings (MH) step to sample tree topology and divergence times $(\cT, \bt)$ (Supplement \ref{supp:inference:mh}), 
(b) a Gibbs sampler with P{\'o}lya-Gamma augmentation plus logistic-distributed auxiliary variables to sample leaf parameters $\bm \eta$ (\ref{supp:algorithm:eta})), 
and 
(c) a Gibbs sampler to sample divergence hyperparemter $c$, diffusion variance $\bsigma^2$, class assignment $\bZ$, and class probability $\bpi$ (Supplement \ref{supp:inference:otherparams}).
We use $\bPhi_{-x}$ to denote all parameters excluding $x$. 

\change{
For Step (a), we follow the MH sampler described in Section 3.1.2 of \cite{yao2022probabilistic}. Given the current tree topology $\cT$ and divergence times $\bt$, a new candidate $(\cT', \bt')$ is sampled from $ p \left( \cT, \bt \mid \bOmega_{-(\cT,\bt)} \right) \propto p(\cT, \bt \mid c) p \left( \bm \eta \mid \cT, \bt, \bsigma^2 \right) $ by randomly detaching a subtree from the current tree and randomly reattaching the subtree back to form a proposed tree. 
We point out that searching in the tree space is not a major challenge for DDT-LCM, even though it is not an easy task in general \citep{yang2000complexity,billera2001geometry}. 
Theoretically, for LCM identifiability, we require the existence of a tri-partition of the $J$ items $[J] = \cup_{m=1}^3 \cA_m$ such that 
$\min(\kappa_1, K) + \min(\kappa_2, K) + \min(\kappa_3, K) \geq 2K+2$ , where $\kappa_m = \prod_{j \in \cA_m} d_j$ \citep[][Theorem 4]{allman2009identifiability}. 
Scientifically, nutrition literature commonly assumes a small $K$ that rarely exceeds 8 to ease interpretability \citep{stephenson2020empirically,park2020latent,uzhova2018generic}. 
Computationally, a relatively small $K$ allows for efficient posterior sampling because the complexity of obtaining one posterior sample is at least $O(K^3)$ for inverting $\bSigma$.  
For a DDT tree with $K$ leaves, the number of unique topologies is $\prod_{k=2}^K (2k-3)$ \citep[][e.g., 3 for $K = 3$ and 105 for $K = 5$]{felsenstein1978NumberEvolutionary}, which is reasonably small for $K \leq 8$.
}

\begin{algorithm}[ht]
    \SetKwInOut{Input}{Input}
    \caption{MH-within-Gibbs Sampler for Posterior Inference}  \label{alg:inference}
    \KwData{Multivariate binary data $\bY$ for $N$ subjects and $J$ food items; food items' group memberships vector $\bM$}
    \Input{the number of classes $K$; initial values of $\cT^{(0)}, \bt^{(0)}, c^{(0)}, \bsigma^{2,(0)}, \bm \eta^{L,(0)}, \bZ^{(0)}, \bpi^{(0)}$; total number of iterations \textit{Niter}; number of burn-ins \textit{Nburn}}
    \For{$r = 1$ to Niter}{
        $(\cT^{(r)}, \bt^{(r)}) \sim \text{MH sampler with acceptance probability in equation \eqref{eq:mh:prob} of the Supplement}$\; %
        \For{$i \in [N], j \in J_g$}{
            $w_{i,j}^{(r)} \sim \text{equation \eqref{eq:cond:w}} $ of the Supplement\; %
            $s_{i,j}^{(r)} \sim \text{equation \eqref{eq:cond:s}} $ of the Supplement\;%
            $\bm \eta_{i,j}^{(r)} \sim \text{equation \eqref{eq:cond:etaL}} $ of the Supplement\;%
        }
        \For{$g \in [G]$}{
            $\sigma_g^{(r)} \sim \text{equation \eqref{eq:cond:sigma}} $\ of the Supplement; %
        }
        \For{$i \in [N]$}{
            $Z_i^{(r)} \sim \text{equation \eqref{eq:cond:z}} $\ of the Supplement; %
        }
        $c^{(r)} \sim \text{equation \eqref{eq:cond:c}} $ of the Supplement %
    }
    \end{algorithm}

We also point out that expecting a small $K$ does not contradict our goal to distinguish weakly separated classes. First, existing nutrition literature presents small $K$ due to the limitations of classical LCMs under weak class separation.
Second, our approach aims to identify meaningfully distinct classes without over-partition (which in the extreme case $K = N$)
Classes are meaningfully distinct if the DPs display discernible variations in food items that explain nutritional differences. Third, as shown in Section \ref{sec:data}, our HCHS/SOL analysis selects $K = 6$, demonstrating a balanced choice.

For Step (b), the posterior distribution of leaf parameter of major food group $g \in [G]$ is
\begin{align} 
    p \left( \bm \eta \mid \bOmega_{-\bm \eta}, \bY \right) &\propto 
    p(\bm \eta \mid \cT, \bt, \bsigma^2) p(\bY \mid \bZ, \bm \eta ) \nonumber \\
    &= \cM \cN_{D_0 \times K} \left( \bm 0, \bOmega, \bSigma \right) 
    \prod\limits_{i=1}^N \prod\limits_{j=1}^{J} 
    \prod_{c=1}^{d_j} \theta_{j,Z_i,c}^{I \{ Y_{i,j} = c\}}. \label{eq:posterior:eta}
\end{align}
To improve estimation accuracy, we draw samples from the exact posterior distribution by introducing two sets of auxiliary variables that follow P{\'o}lya-Gamma distributions and logistic distributions in the Gibbs sampler \citep{dalla2021polya}.

For Step (c), we derive the full conditional distribution for each variable and apply a Gibbs sampler. We point out that estimating the divergence function hyperparameter $c$ in DDT-LCM is challenging, due to the typically small number of leaves $K$ in dietary pattern analysis (see equation \eqref{eq:cond:c} of the Supplement). %
However, since our primary goal is to improve pattern estimation by incorporating a tree prior, estimating $c$ is not a major concern. This contrasts with \cite{yao2022probabilistic} where a larger tree with 20 cancer drugs as leaves allows for more precise inference of $c$.

\paragraph*{Posterior Summaries} \label{sec:posterior:summary}
We propose the following strategy for posterior estimation and inference for the tree parameters $(\cT, \bt, \bsigma^2, c)$ and the LCM parameters $(\bTheta, \bZ, \bpi)$. Among the tree parameters, the tree structure $(\cT, \bt)$ is obtained via the maximal clade credibility (MCC) tree, whose topology achieves the maximum product of posterior clade probabilities and branch lengths are set the median branch lengths for each of the clades \citep{heled2013LookingTrees}. 
Point and interval estimates for the divergence hyperparameter $c$ and diffusion variance $\bsigma^2$ are obtained by posterior means and credible intervals. The LCM parameters are computed using the posterior means and credible intervals for the item response probabilities $\bTheta$ and class probability $\pi$. Individual memberships $\bZ$ are assigned as the class with highest posterior probability of class assignment.

The above algorithm and summaries are available in the R package \href{https://cran.r-project.org/web/packages/ddtlcm/index.html}{\texttt{ddtlcm}}.


\section{Simulation} \label{sec:simulation}
\change{
We performed two sets of numerical experiments. The first set used fully synthetic data to demonstrate model performances under four trees with decreasing class separation; see Section \ref{sec:sim:synthetic}. The second set mimicked our motivating data in terms of sample size and degree of class separation to demonstrate the effectiveness for the proposed DDT-LCM with a data-driven choice of $K$; see Section \ref{sec:sim:real}. 

}

The performance of our model was compared with the following measures: accuracy in estimating the class profiles, individual class assignments, and tree structures. 
\change{We compared our model with five baseline methods, among which three were variants of the proposed DDT-LCM, and two were based on the commonly used classical Bayesian LCM with independent priors. }
We list all compared models below:
(i)``DDT-LCM'': the proposed model where the tree is unknown and to be estimated; 
(ii)``DDT-LCM (true tree)'': DDT-LCM with the tree fixed at the true tree structure, (omitting Algorithm \ref{alg:inference}, line 2); 
(iii) ``DDT-LCM (misspecified tree)'': DDT-LCM with the tree fixed at a misspecified structure; 
(iv)``DDT-LCM (homogeneous var)'': DDT-LCM without group-specific variances; 
(v) ``BayesLCM (heterogeneous var) + HC'': Bayesian LCM and independent normal priors on the logit-transformed patterns with group-specific variance parameters $\sigma_g^2$, followed by agglomerative hierarchical clustering (HC) on the estimated patterns to obtain a tree over the classes; 
(vi) ``BayesLCM (homogeneous var) + HC'': the same as model (v) except with a homogeneous variance parameter ($\sigma^2_g = \sigma^2)$. 
\change{Method (ii) is designed to show gain in estimation accuracy if the true between-class dependence structure is known. Methods (iii) and (v) are to demonstrate the loss in estimation accuracy if the between-class dependence is specified arbitrarily. Methods (iv) and (vi) demonstrate the need for flexibe food group-specific variances.}

\sloppy Three performance metrics were evaluated. First, given the one-to-one correspondence between trees and covariance matrices (Proposition \ref{prop:marginalleaf}), we assessed tree recovery accuracy by computing the Frobenius norm of the difference between estimated $\hat{\bSigma}$ associated with the maximum clade credibility (MCC) tree and the truth: $F \left( \hat{\bSigma} \right) = \left \lVert \hat{\bSigma} - \bSigma \right\rVert_F = \left[ \sum_{k=1}^K \sum_{l=1}^K ( \hat{\bSigma}_{k,l} - \bSigma_{k,l} )^2 \right]^{1/2}$. 
Second, we calculated the root mean squared errors (RMSEs) for the estimated conditional probabilities: $\mathrm{RMSE} \left( \hat{\bTheta} \right) = \left[ (KJ)^{-1} \sum\limits_{k=1}^K \sum\limits_{g = 1}^G \sum\limits_{j = 1}^{J_g} \left( \hat{\theta}_{g,j,k} - \theta_{g,j,k} \right)^{2} \right]^{1/2}$. 
Third, we measured the agreement between true and estimated individual class assignments by the adjusted Rand index \citep[ARI,][]{hubert1985comparing}, a chance-corrected measure between $-1$ and $1$ with values near $1$ indicating stronger agreement. 
Additionally, we evaluated the empirical coverage probability of the 95\% credible intervals for diffusion variance $\bsigma^2$ (see Section \ref{supp:sec:simulation:syntheticsetup} of the Supplement). %

\subsection{Simulation I: Synthetic Data} \label{sec:sim:synthetic}

We considered LCMs with $K = 3$ and $J = 80$ items categorized into $G = 7$ major food groups with $J_1 = \cdots = J_5 = 10$ and $J_6 = J_7 = 15$. We assumed that all items have binary consumption with $d_j = 2 \ \forall j \in [J]$. Four distinct tree structures were considered corresponding to different levels of class separation. %
Trees 1 and 2 represent strong and moderate class separation. Trees 3 and 4 represent weak class separation.
For each tree, we simulated 100 independent data sets for $N \in \{100, 200, 400\}$ individuals. See Section \ref{supp:sec:simulation:syntheticsetup} of the Supplement
for additional setup details.


Figure \ref{fig:sim_boxplot_combined} presents model performance results. 
In the top row, DDT-LCM achieves similar or superior recovery of the true tree structure compared to models (v) and (vi). Note that the Frobenius norms of models (ii) and (iii) are constant because their tree structures are fixed during estimation. As classes separation decreases from Tree 1 to Tree 4, all methods exhibit higher errors in tree recovery, but DDT-LCM remains more accurate and stable. This is expected because (v) and (vi) perform LCM estimation and post hoc hierarchical clustering in separately, failing to propagate uncertainty into tree estimation. In contrast, DDT-LCM jointly estimates the tree and other model parameters, fully acconuting for uncertainty. 

DDT-LCM is comparable to model (ii) true tree case and outperforms the other models
with lower RMSEs of the estimated conditional probabilities (middle row, Figure \ref{fig:sim_boxplot_combined}) and higher ARIs of individual class memberships (bottom row, Figure \ref{fig:sim_boxplot_combined}). This indicates that jointly learning between-class similarity and the guiding tree improves parameter estimation. If the true tree structure is known, additional accuracy gain is achieved. On the other hand, if the tree is misspecified at a structure far from the truth, estimation accuracy may be much worse than DDT-LCM and no better than the plain BayesLCM, which is particularly problematic under smaller sample sizes (e.g., $N=100$). 
DDT-LCM becomes more advantageous to other methods for weaker separation between the classes (from Tree 1 to 4). Larger sample sizes tend to make up for the performance disadvantage of other methods ($N=100$ to $400$). In particular, in Tree 1 with well-separated classes, DDT-LCM performs similarly to the competing methods in all metrics. In this case, BayesLCM may be more computationally efficient and practically applicable because it does not require sampling for the trees. As the classes become less separated, DDT-LCM takes advantage of between-class similarity and leads to accuracy gain. When sample sizes are small, information from tree-guided between-class similarity dominates the posterior distribution of the parameters. This dominance diminishes when a large sample ($N = 400$) provides sufficient information in this small-scale fully synthetic study.

The better estimation of the LCM parameters using model (i) than model (iv) and (v) than (vi) implies that group-specific variances cannot be ignored. The coverage probabilities of the 95\% credible intervals of $\bsigma^2$ are close to the nominal level for larger sample sizes (Figure \ref{fig:sim_K3_sigma}(a) 
of the Supplement). The posterior mean estimates of the last two groups $\sigma_6^2$ and $\sigma_7^2$ have higher levels of uncertainty than the first five groups (Figure \ref{fig:sim_K3_sigma}(b)
of the Supplement), because the conditional probabilities of items in the last two groups are closer to the boundaries 0 or 1, making estimation more challenging. In summary, we demonstrate that by leveraging similarity information shared between patterns and accounting for varying levels of class separation by major food groups, DDT-LCM improves accuracy relative to common alternatives in estimating the tree structure, conditional probabilities, and class assignments.

\begin{figure}[h!]
    \centering
    \begin{minipage}[c]{\linewidth}
        \centering
        \includegraphics[width=\linewidth]{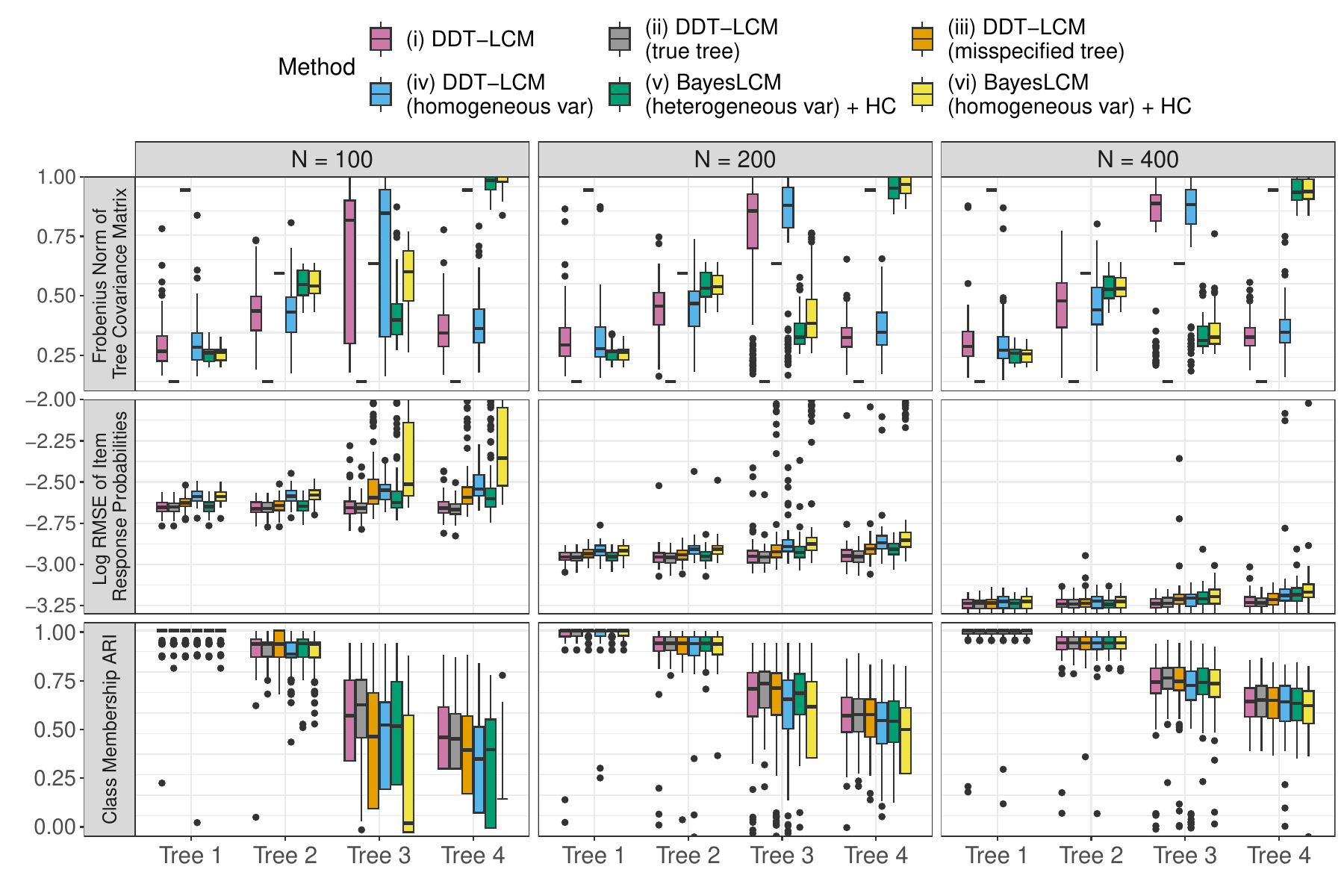}
    \end{minipage}
    \vspace{-5mm}
    \caption{Simulation I: fully-synthetic data to compare parameter recovery performance for different trees and sample sizes for {\it top}) tree structure, \textit{middle}) class profiles, and \textit{bottom}) class memberships. \textit{Middle/Bottom)}: left to right: (i) -- (vi). \textit{Top}): (ii) and (iii) are omitted. Single linkage was used in (v) and (vi); results are similar for other linkage functions.}
    \label{fig:sim_boxplot_combined}
\end{figure}

\subsection{Simulation II: Semi-Synthetic Data} \label{sec:sim:real}

We mimicked the real data in the HCHS/SOL study to investigate whether DDT-LCM can confer statistical benefits under the realistic sample size and between-class separation observed in the data. We also sought to evaluate a method to perform data-driven selection of $K$ in such scenarios. To this end, we simulated $J = 78$ granular items categorized into $G = 7$ major food groups, and $N = 400, 800$ subjects in $K = 6$ latent classes. See Section \ref{sec:sim:real:setup} of the Supplement for the detailed setup. %

\begin{figure}[ht]
    \centering
    \begin{minipage}[c]{0.75\linewidth}
        \centering
        \includegraphics[width=\linewidth]{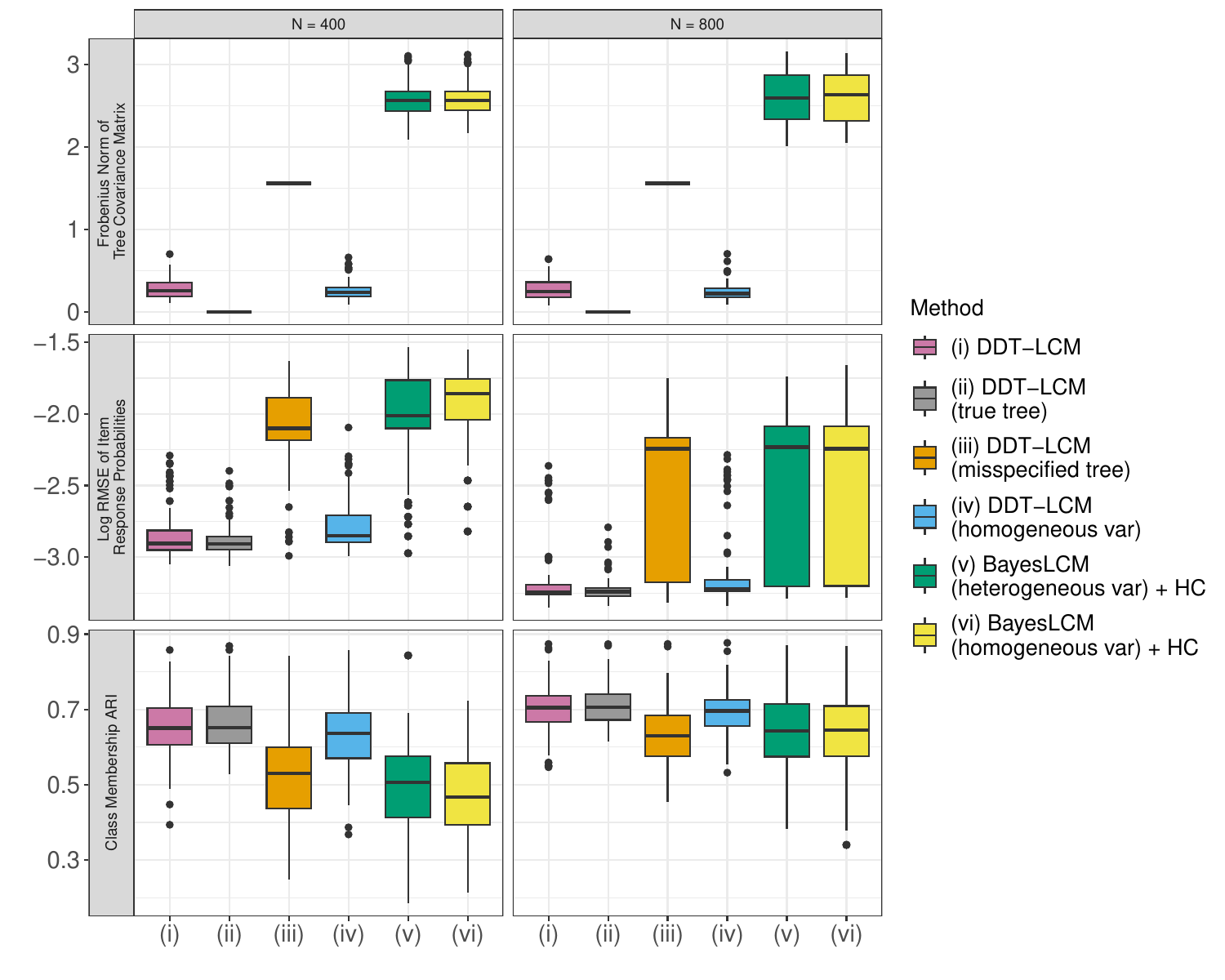}
    \end{minipage}
    \vspace{-5mm}
    \caption{Simulation II to assess parameter recovery performance under weak separation and different sample sizes that mimic the real data: 
    {\it top}) tree structure, \textit{middle}) response probabilities, and \textit{bottom}) class memberships.  For methods (v) and (vi), single linkage is used; results are similar for other linkage functions.
    }\label{fig:sim_boxplot_real_combined}
\end{figure}

Figure \ref{fig:sim_boxplot_real_combined} displays the estimation results of DDT-LCM compared to alternative methods, assuming that the true $K$ is known. Under the realistic weak class separation scenario, DDT-LCM is capable of more accurately recovering the true tree and LCM parameters under sample sizes below and above our real data size ($N = 496$), suggesting its practical usefulness in deriving patterns for small-sized subpopulations. Moreover, we notice that for the majority of the simulated datasets, BayesLCM (with either heterogeneous or homogeneous variance) tends to merge two or more latent classes by forcing the class probabilities of these classes to near zero. 
The independent priors on $\bTheta$ do not facilitate BayesLCM to leverage similarity information across classes, leading to incorrect merging of classes. These results demonstrate the ability of DDT-LCM to handle complicated weakly separated real data scenarios by producing reliable estimates. Figure \ref{fig:sigma_llk_difference}
of the Supplement 
indicates that the proposed model (i) produces higher predictive log-likelihoods than model (iv). Section \ref{sec:simulation:real:variance}  of the Supplement 
includes a discussion about diffusion variance parameter estimation. 

The performance of DDT-LCM in the simulations has been demonstrated under a known $K$. To provide a practical estimation pipeline applicable to real-world data, we can select $K$ in a data-driven manner. Our rationale here is to apply a practically useful criterion that leans towards a model with good out-of-sample predictive performance while remaining parsimonious. 
Viewing $K$ as a hyperparameter, we choose the $K$ that yields the average largest predictive log-likelihood on validation datasets from 5-fold cross-validation. 
Specifically, the observations of $N$ individuals are randomly split into a training set and a testing set according to a 4:1 ratio. For the $s$-th training set, $s \in [5]$, we apply the Gibbs sampler and obtain the posterior means of the class prevalences $\hat{\pi}$ and response probabilities $\hat{\theta}$. The predictive likelihood on the corresponding testing set is computed as 
\begin{equation} \label{eq:crossvalidation}
    p_{K,s}^{test} = \prod\limits_{i \in I^{test}_s} \left[ \sum\limits_{k=1}^K \hat{\pi}_k \prod\limits_{g=1}^G \prod\limits_{j=1}^{J_g} \left( \hat{\theta}_{g,j,k} \right)^{Y_{i,g,j}} \left( 1 - \hat{\theta}_{g,j,k} \right)^{1 - Y_{i,g,j}} \right],
\end{equation}
where $I^{test}_s$ denotes the indices of individuals belonging to the $s$-th testing set. The average predictive log-likelihood is then calculated as $l_{K}^{test} = \frac{1}{5} \sum\limits_{s=1}^5 \log p_{K,s}^{test}$. The model with a larger $l_{K}^{test}$ is preferred. 
We illustrate the criterion using the semi-synthetic data and results can be found in Section \ref{sec:sim:chooseK} of the Supplement. 


\section{Application to Dietary Intake Data} \label{sec:data}

\subsection{Data and Method} \label{sec:data:background}
The Hispanic Community Health Study/Study of Latinos (HCHS/SOL) is a multi-center, community based cohort study of Hispanic/Latino adults in the United States \citep{lavange2010SampleDesign}. 
In this analysis, we focus on dietary habits of $N = 496$ participants with South American ethnic background.
We selected this subgroup for its smaller sample size compared to other ethnic backgrounds and its known within-subgroup diet heterogeneity.
Dietary intake was obtained from study participants via two 24-hour dietary recalls collected at baseline (2007-2011). These recalls were conducted using the Nutrition Data System for Research (NDSR) software developed by the Nutrition Coordinating Center at the University of Minnesota. Foods were summarized into 7 broad food groups created from 157 NDSR food codes. 
Intake was quantified by servings per day. Participants with at least one reliable recall, defined by HCHS/SOL staff, were included for analysis. Participants with more than one reliable recall were averaged over the two days recorded. To best capture the heterogeneous variation of foods consumed, 79 foods were excluded due to extremely high ($>97.5\%$ exposure presence) or low ($< 2.5\%$ exposure presence).

Since most foods were consumed for a low amount, our analysis considers dichotomized participant responses, where $Y_{ij} = 1$ or 0 denotes presence or absence of exposure to food item $j$. 
Responses to $J = 78$ food items were curated, which belong to $G = 7$ major nutrition groups: fat, fruit, grain, meat, dairy, sugar, and vegetables. 
A detailed list of food items and their major food groups is provided in Table \ref{tab:item_labels} of the Supplement. %
We applied the proposed DDT-LCM to the cohort for analysis. 
The candidate values for $K$ include $\{3,4,5,6,7,8\}$ and we would choose the $K$ that produced the largest average predictive log likelihood via five-fold cross-validation (see Section \ref{sec:sim:chooseK} of the Supplement).
For the optimal $K$, we ran the Gibbs sampler for 12,000 iterations and discarded the first 7,000 samples as burn-ins.

\subsection{Results} \label{sec:data:results}	
\begin{sidewaysfigure}
    \centering
    \includegraphics[width=\linewidth]{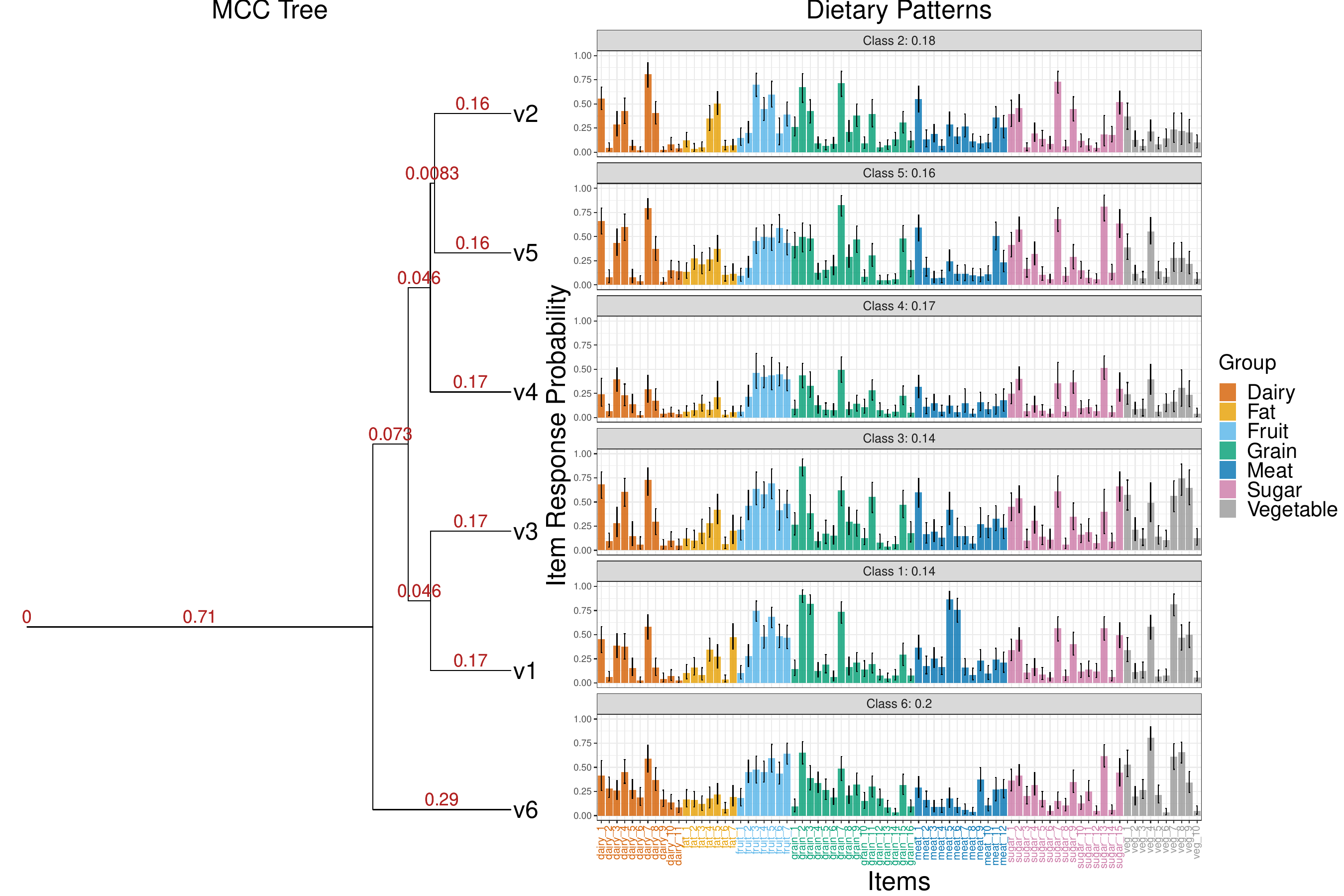}
    \caption{HCHS/SOL result with $K = 6$ latent classes using binary responses from $N = 496$ US adults with South American ethnicity. Left: the MCC tree structure over the latent classes. Numbers indicate the corresponding branch lengths. Right: dietary profiles for the $J = 78$ food items belonging to $G = 7$ different pre-defined nutrition groups, distinguished by different colors. The description of individual items is provided in Table \ref{tab:item_labels} of the Supplement 
    with matching item labels on the x-axis. The numbers after the class labels in the facets indicate class prevalence. Error bars show the 95\% credible intervals from the posterior distribution.
    }
    \label{fig:data_result}
\end{sidewaysfigure}

Based on our model selection criteria, we selected $K=6$ classes. We observed good convergence and mixing of our sampling algorithm. Figure \ref{fig:data_result} displays the derived dietary patterns and MCC tree under the selected model. The estimated diffusion variances for the 7 major food groups are 3.64 (fat), 1.75 (fruit), 2.88 (grain), 3.67 (meat), 3.58 (dairy), 2.60 (sugar), and 3.62 (vegetable). These estimates are consistent with the dietary pattern shown in Figure \ref{fig:data_result}. The smallest variance occurs amongst fruit items. The probabilities of exposure to these fruit items do not vary much across latent classes, but the largest source of variation is identified in class 2. Fat, meat, dairy, and vegetable groups display the largest degrees of variability, implying that the dietary patterns exhibit major differences in items belonging to these major food groups.
The average log-likelihoods in five-fold cross-validation were $-3738.771$ and $-3746.070$ for DDT-LCM and ``DDT-LCM (homogeneous var)'', respectively, indicating that DDT-LCM with group-specific variance parameters produced better predictive performance.


Classes 1 and 2 shared slightly similar behaviors in fruit, sugar, and vegetable food groups, and have a cophenetic distance of 0.834 on the MCC tree. Class 1 had higher probabilities of exposure to refined grain dry mixes, lean poultry, fried chicken, unsweetened coffee substitutes, white and fried potatoes, and other vegetables. Class 2 had higher probabilities of exposure to reduced fat salad dressing, cereal and bread with some whole grain, and whole grain snacks, lamb, milk and flavored milk, and all vegetables except for pickled foods. Class 3 shared some of these similarities with class 1 and class 2 (with a cophenetic distance of 0.815), but differed significantly in patterns within fat and meat groups with much lower exposure probabilities. Higher probabilities of exposure of this class were found in fruit-based snack, fried shellfish, sweetened fat free yogurt, unsweetened coffee substitutes, white potatoes and fried vegetables. Class 4 shared slightly similar dietary behaviors in all food groups with class 5 (with a cophenetic distance of 0.877) and had higher probabilities of exposure to a number of items,
including salad dressing, citrus fruit, avocado and similar, dry grain mixes, crackers, lamb, lean poultry, whole milk, yogurt, sweetened fruit drinks, dessert, and dark-green vegetables. Class 5 had much higher probabilities of exposure to reduced-fat flavored milk, unsweetened coffee substitutes, and white potatoes. Class 6 shared common group-level attributes with class 4 (grains, meat, dairy, sugar) and class 5 (fruit, vegetables), at a cophenetic distance of 0.861. In the dairy group where classes 4 and 6 shared similarities, class 6 exhibited slightly higher probability of exposure to flavored milk.

\change{
Further post-hoc analysis on the relationship between subject characteristics and the DDT-LCM derived dietary patterns can shed light on nutritional implications. We compared individual characteristics among subjects with relatively healthier dietary pattern (class 2) and less healthy pattern (class 1). We find that the proportion of individuals aged between 18 and 44 years (as opposed to over 45 years) is slightly higher in class 2 (63\%) than class 1 (51\%). Additionally, class 2 consists of 54\% females whereas class 1 consists of 48\% females. This suggests that the subgroup of young females are mroe likely to commit a healthier pattern. This finding is also consistent with existing nutrition studies \citep{maldonado2021dietary}.
}

We also compared these results with the standard Bayesian LCM, fit under $K = 3, 4, 5, 6, 7$ classes. With the exception of the 3-class model, Bayesian LCMs (with either homogeneous or heterogeneous variances) did not converge and nearly zero class probabilities were estimated in at least one latent class.
The item probability estimates in the sparse classes were all close to 0.5 with 95\% credible intervals covering the complete probability range between 0 and 1. 
Although we might discard the sparse classes after implementing BayesLCM, the number of non-sparse classes was not consistently produced by the model under different $K$ values. This phenomenon echoes findings in prior literature \citep{lubke2007PerformanceFactor,park2018RecommendationsSample,weller2020LatentClass} that when latent classes are not sufficiently separated, convergence failure likely occurs due to small sample sizes.

Comparing with the dietary patterns estimated using the standard Bayesian LCM with 
$K = 3$ displayed in Figure \ref{fig:data_result:bayeslcm} %
of the Supplement, we see that DDT-LCM with $K = 6$ is capable of detecting nuanced differences in food exposure by separating out more dietary patterns. In particular, the standard Bayesian LCM estimates the probabilities of exposure to low-fat yogurt to range between 0.5 and 0.65 for all three dietary patterns, while DDT-LCM considers these probabilities to be 0.78 (class 2), 0.28 (class 4) and 0.76 (class 5) in three additional patterns. Similarly, in addition to the patterns of vegetable items estimated by the standard Bayesian LCM, DDT-LCM also identifies class 1 to have very high probability (0.8) of exposure to other vegetables while much lower probabilities of exposure to fried vegetables and vegetable juice.

Additionally, the selected number of classes $K=6$ resides on the upper end of typically selected numbers (2 to 6) in existing nutrition literature \citep{park2020latent,dalrymple2023evaluation,dalmartello2020dietary,stephenson2023identifying} that applies the standard LCM. This is not surprising because when between-class similarities are not recognized, the standard LCM fails to separate closely related classes and thus favors a smaller $K$.

\section{Discussion} \label{sec:discussion}
	
Using LCMs, we derived dietary patterns of a small subpopulation of HCHS/SOL participants with South American ethnic background. 
We enhanced the inference of dietary patterns by introducing a tree-regularized Bayesian LCM that infers a hierarchical relationship among dietary patterns to facilitate sharing of statistical strength to make better estimates using limited data. Simulation and data analysis demonstrated that our method improved estimation of dietary patterns and individual class assignments relative to existing techniques based on classical Bayesian LCMs. 

It is worth noting that DDT-LCM is perfectly suitable under larger sample sizes and produces comparable performances as classical LCMs at additional computational expense. In practice, because the boundary between large and small sample sizes is often unclear and must be determined relative to the actual degree of separation between dietary patterns in the data, DDT-LCM can guard against potential numerical and statistical instability.

\change{
DDT-LCM derives dietary patterns as an exposure analysis among a subgroup that is typically undersized compared to the rest of the HCHS/SOL cohort. 
This approach enables the study of other small subpopulations, such as San Diego participants of Central American background or Bronx participants of Cuban background, which were excluded from analyses in\cite{stephenson2020empirically} and \cite{vito2022SharedEthnic}. 
Unlike RPC, we did not stratify further by study site.
DDT-LCM also differs from RPC in application contexts. DDT-LCM is tailored to a single small subgroup, whereas RPC identifies global and local differences across multiple pre-specified subpopulations. 
If coupled with DDT-LCM, RPC may become more sensitive at identifying the subtle local differences in dietary patterns.

There are additional limitations of the current study. We did not consider measurement errors in the dietary recall data in our application. For example, memory biases may lead to reporting food items that are actually not consumed. A relevant discussion about correcting for measurement errors is included in Section \ref{sec:measurementerror}
of the Supplement. Moreover, DDT-LCM assumes the food item responses are independent within a dietary pattern, as mentioned in Remark \ref{remark:dependency}.  
For example, whole milk and low-fat milk (both belong to the diary group) may be substitutes because having one could decrease the chance of having the other. An important extension of the proposed model is to incorporate such nutrition knowledge-informed dependency such that nutritionally distinct dietary patterns can be derived.
The dietary recalls in the HCHS/SOL were collected only at baseline and thus our analysis was cross-sectional. Other studies have considered longitudinal designs to collect diet data \citep{nouri2021LongitudinalAssociation,aljahdali2022SedentaryPatterns}. 
Extension to longitudinal settings may better investigate how diet impacts health outcomes, especially in small subpopulations.
}

\section*{Data Availability and Code}

\sloppy 
The data that support the findings of this study are protected under a Data and Materials Distribution Agreement (DMDA). Access to the application data is available upon request from the \href{https://sites.cscc.unc.edu/hchs/}{study website}.
Code to reproduce simulations is available online at \href{http://github.com/limengbinggz/ddtlcm}{\nolinkurl{http://github.com/limengbinggz/ddtlcm}}.

\begin{acks}[Acknowledgments]  \label{sec:Acknowledgement}
The authors gratefully acknowledge Daniela Sotres-Alvarez and Anna-Maria Siega-Riz for helpful comments and feedback on earlier drafts of this work. The authors also thank Tsung-Hung Yao for sharing code to implement the MH algorithm. 
\end{acks}

\begin{funding}
The research is partially funded by Michigan Institute for Data Science (MIDAS) to Mengbing Li and Zhenke Wu. 
The HCHS/SOL is a collaborative study supported by contracts from the National Heart, Lung, and Blood Institute (NHLBI) to the University of North Carolina (HHSN268201300001I / N01-HC-65233), University of Miami (HHSN268201300004I / N01-HC-65234), Albert Einstein College of Medicine (HHSN268201300002I / N01-HC-65235), University of Illinois at Chicago – HHSN268201300003I / N01-HC-65236 Northwestern University), and San Diego State University (HHSN268201300005I / N01-HC-65237). The following Institutes/Centers/Offices have contributed to the HCHS/SOL through a transfer of funds to the NHLBI: National Institute on Minority Health and Health Disparities, National Institute on Deafness and Other Communication Disorders, National Institute of Dental and Craniofacial Research, National Institute of Diabetes and Digestive and Kidney Diseases, National Institute of Neurological Disorders and Stroke, NIH Institution-Office of Dietary Supplements.
\end{funding}


\begin{supplement}
\textbf{Supplement \ref{supp:sec:ddt}: Generative Process of the Dirichlet Diffusion Tree.} %

\textbf{Supplement \ref{sec:matrixlogitnormal}: Matrix Logistic Normal Distribution.} %

\textbf{Supplement \ref{supp:sec:marginalprior}: Marginal Prior with Closed-Form Likelihood.} %

\textbf{Supplement \ref{supp:sec:treecov_example}: The Tree-Structured Covariance Matrix.} %

\textbf{Supplement \ref{supp:sec:inference}: Posterior Sampling Algorithm.} %

\textbf{Supplement \ref{supp:sec:simulation}: Additional Simulation Study Details.}%

\textbf{Supplement \ref{supp:item_list}: Additional Results in HCHS/SOL Data Analysis.} %

\textbf{Supplement \ref{sec:multigroup}: Potential Extension to Multiple Groups.} %

\textbf{Supplement \ref{sec:measurementerror}: Discussion about Measurement Error in Dietary Surveys.} %
\end{supplement}


\bibliographystyle{imsart-nameyear} 
\bibliography{references}       

\newpage
\appendix

\title{Supplement to ``Tree-Regularized Bayesian Latent Class Analysis for Improving Weakly Separated Dietary Pattern Subtyping in Small-Sized Subpopulations''}
\runtitle{Tree-Regularized Bayesian Latent Class Analysis}


\renewcommand{\thesection}{S\arabic{section}}
\renewcommand\thefigure{\thesection.\arabic{figure}}    
\renewcommand\thetable{\thesection.\arabic{table}}    
\numberwithin{equation}{section}
\makeatletter 
\renewcommand\theequation{\thesection.\arabic{equation}}    
\newcommand{\section@cntformat}{Supplement \thesection:\ }
\makeatother

\blfootnote{\textsuperscript{$\ast$}Co-senior authors. }

\section{Generative Process of the Dirichlet Diffusion Tree} \label{supp:sec:ddt}

The DDT process provides a family of nonparametric priors for distributions over exchangeable random quantities that arise from a latent branching process. DDT processes generalize Dirichlet processes by capturing the hierarchical structure present in complex distributions by means of a latent diffusion tree. A joint distribution is specified on rooted binary trees with a fixed number of leaves and Gaussian-distributed node parameters. It is also generalizable to fit non-Gaussian distributions through transformation \citep{knowles2015pitman}. 


\change{
Here we consider a DDT process that generates a $K$-leaf rooted binary tree with latent states in the space $\RR^{D_0}$. 
We label the $K$ leaf nodes as $1, \ldots, K$, the root node as $K+1$, and the $K-1$ internal nodes as $K+2, \ldots, 2K$. 
In essence, the $K$ leaf latent states are the final destinations of $K$ different ``particles" traveling from the origin $\bm 0 \in \RR^{D_0}$ at time $t = 0$ until $t = 1$ according to a multivariate Brownian motion in a self-reinforcing scheme. 
Similar to the Chinese restaurant process \citep{aldous1985exchangeability}, the self-reinforcing scheme specifies a branching process where the more particles follow a particular path, the less likely subsequent particles will diverge off this path. 
Specifically, the first particle simply travels without divergence according to a Brownian motion originating at the root latent state $\bm \eta_{K+1} = 0$ at time $t = 0$, and the first leaf latent state $\bm \eta_1$ is obtained as the particle stops at $t = 1$. 
The second particle starts at the origin again and follows the path of the first particle until some divergence time $t$, after which it travels according to an independent Brownian motion and results in the second leaf latent state $\bm \eta_2$. 
The instantaneous probability of diverging on the infinitesimal interval $[t, t + dt)$ is $\frac{a(t) dt }{ m }$, where $m$ denotes the number of particles that have previously traversed the current path (so $m = 1$ for the second particle). 
Here, $a(t)$ is a divergence function satisfying $\int_0^1 a(t) dt = \infty$ such that each particle branches before $t = 1$ almost surely. 
Similarly, each of the remaining particles follows an existing path initially. If a particle does not diverge before reaching a previous divergence point, it will follow one of the existing paths with probability proportional to the number of particles that previously travel along each path. 
After generating $K$ particles, we obtain tree topology $\cT$, branch times $\bt$, latent states at internal and leaf nodes. An graphical illustration of the above diffusion dynamics can be found in Figure 2(A) of \cite{yao2022probabilistic}.
}

\change{
In the following, we describe how the divergence function $a(t)$ impacts the branching process, which is also essential for deriving the posterior distribution of DDT-LCM. 
For a particle currently traveling along an existing path between $[t_a, t_b]$ that has previously been visited by $m$ particles, the likelihood of the particle diverging at time $t > t_a$ is 
$ P(\text{branch in } [t_a, t]) = 1 - \exp \{ - [A(t) - A(t_a)] / m \}, $
where $A(t) = \int_0^t a(s) ds$ is called the cumulative branching function.
For the choice of $a(t) = c / (1-t)$ in \eqref{eq:divfunction} of the main paper, we have $A(t) = -c \log(1-t)$ and hence $P(\text{branch in } [t_a, t]) = 1 - \left( \frac{1 - t}{1 - t_a} \right)^{c/m}$. 
Therefore, compared to a smaller value of $c$, a larger $c$ places higher probability on large $t$, resulting in later divergence time and thus stronger prior dependence between different leaf parameters. 
}

DDT jointly controls the prior on tree topology and branch lengths through athe divergence function $a(t)$. This joint prior provides more flexibility than the commonly used uniform prior on tree topology and i.i.d. exponential or uniform priors on branch lengths \citep{huelsenbeck2001mrbayes}, which are typically employed for ease of computation. In addition, the i.i.d. priors can cause overly long tree when the latent classes are highly similar \citep{brown2010trees}.

\section{Matrix Logistic Normal Distribution} \label{sec:matrixlogitnormal}

Suppose that $\bV \sim \mathrm{MLN}(\bM, \bU, \bW)$, as defined in Definition \ref{def:matrix:logitnormal}. 
The probability density function of $\bV$ is
\begin{align} \label{eq:matrix:logitnormal}
f_{\bV} (\bv; \bM, \bU, \bW) =
\dfrac{ 
\exp \left\{ 
    -\frac{1}{2} \mathrm{Tr}
    \left( \bW^{-1} 
    \left( \bv^* - \bM \right)^\top 
    \bU^{-1} 
    \left( \bv^* - \bM \right) 
    \right)
    \right\}
}{
(2 \pi)^{K(D-1)/2} 
|\bW|^{(D-1)/2} 
|\bU|^{K/2} 
\prod\limits_{k=1}^{K}
\prod\limits_{d=1}^{D} v_{d,k}
},
\end{align}
where $\bv^* = \left( v_{d,k}^* \right)_{(D-1) \times K}$ with log odds $v_{d,k}^* = \log \frac{v_{d,k}}{ v_{D,k} }, d \in [D-1], k \in [K]$, and Tr denotes the trace of a matrix.

\section{Marginal Prior with Closed-Form Likelihood} \label{supp:sec:marginalprior}
We briefly describe the joint density of $(\bm \Theta, \cT, \bt)$ marginalized over all intermediate stochastic paths between tree nodes.
For an internal node $k \in \{K+2, \ldots, 2K\}$, let $l(k)$ and $r(k)$ be the number of leaf nodes under the left and right child of $k$, respectively, and let the total number of leaf nodes under node $k$ be $m(k) = l(k) + r(k)$. 
The structure of a tree (topology and branch lengths) generated under equation \eqref{eq:ddt:prior} of the main paper can be viewed as a set of segments $\cS(\cT) = \left\{ [kh]: 0 < t_k < t_h < 1, k, h \in \{K+1, \ldots, 2K\} \right\}$. 
The joint probability density of leaf latent states and tree structure, conditional on the diffusion variance $\bsigma^2$ and the divergence function $a(t) = c / (1-t)$, is given by
\begin{align}
    \begin{split}\label{eq:ddt:jointdensity}
        & \ \ \ \ \ \ \ P \left( \bm \Theta, \cT, \bt \mid c, \bsigma^2 \right) = \\
        &\ \left[ \prod\limits_{[kh] \in \cS(\cT)}  \right.
        \underbrace{\vphantom{\prod\limits_{j=1}^{J}} \frac{ (l(h) - 1)! (r(h) - 1)! }{ (m(h) - 1)! } }_{\substack{\text{tree topology } \\ P( [kh] )}}
        \underbrace{ \vphantom{\prod\limits_{j=1}^{J}} c (1 - t_h)^{ c J_{h} - 1} }_{\substack{\text{branch lengths } \\ P(t_h \mid [kh], c)}} 
        \left. \vphantom{\prod\limits_{j=1}^{J}} \right]
        \underbrace{ \vphantom{\prod\limits_{j=1}^{J}} 
        \prod\limits_{j=1}^{J} \mathrm{MLN}_{d_j \times K} \left( \bm 0, \sigma_{r_j}^2 I_{d_j-1}, \bSigma \right)}_{
        \vphantom{\prod\limits_{j=1}^{J}} \text{patterns } P(\bTheta \mid \cT, \bt, \bsigma^2)
    }
    \end{split}
\end{align}
where $J_{v} = H_{m(v)-1} - H_{l(v)-1} - H_{r(v)-1}$ and $H_{n} = \sum\limits_{i=1}^n 1/i$ is the $n$-th harmonic number. 
Refer to \cite{knowles2015pitman} for a detailed derivation of the joint density.

\section{The Tree-Structured Covariance Matrix} \label{supp:sec:treecov_example}

The tree-structured covariance matrix $\bSigma$ not only has an important role of characterizing the DDT tree in this paper, but also sees important applications in  phylogenetic tree literature \citep{revell2008phylogenetic,yao2022probabilistic}.
In fact, the tree-structured covariance matrix in our DDT context always has 1's on the diagonal, and is \textit{strictly ultrametric} and therefore nondegenerate almost surely \citep{martinez1994inverse,nabben1995generalized}. Any strictly ultrametric matrix has one-to-one correspondence to a rooted binary tree \citep{martinez1994inverse}, and this correspondence can be easily extended to an improper rooted binary tree (i.e., a DDT tree) by subtracting the length of the branch attached to the root node from all elements in $\bSigma$. As a result, the distance between the associated matrices of two DDT trees is a reasonable metric for comparing the trees structures (Section  \ref{sec:simulation} 
of the main paper). 

\begin{figure}[h!]
    \centering
    \includegraphics[width=0.35\textwidth]{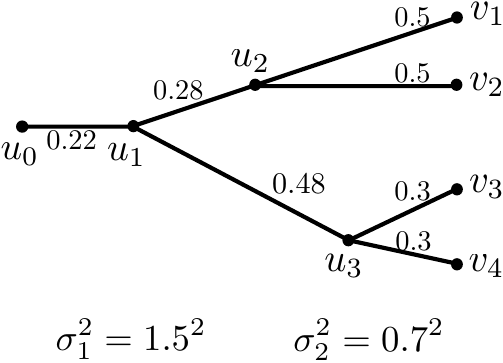}
    \caption{Example of a tree structure. For illustration purpose, we assume two groups of items are present with different diffusion variance parameters.}
    \label{fig:example_covariance}
\end{figure}

Suppose we have items categorized into $G = 2$ groups. The tree in Figure \ref{fig:example_covariance} in the main paper is a possible structure over a 4-class LCM. The resulting tree-structured covariance matrix is
\begin{equation*}
    \bSigma = \begin{pmatrix}
        1 & 0.5 & 0.22 & 0.22 \\
        0.5 & 1 & 0.22 & 0.22 \\
        0.22 & 0.22 & 1 & 0.7 \\
        0.22 & 0.22 & 0.7 & 1
    \end{pmatrix}.
\end{equation*}
The diffusion variance of groups 1 and 2 are $\sigma_1^2 = 1.5^2$ and $\sigma_2^2 = 0.7^2$, respectively. Therefore, the row covariance of the two groups, as defined in equation (5) 
of the main paper, are $1.5^2 \bSigma$ and $0.7^2 \bSigma$, respectively. The cophenetic distance between $v_1$ and $v_2$ is 0.5, and that between $v_1$ and $v_3$ is 0.78.

\section{Posterior Sampling Algorithm} \label{supp:sec:inference}
We provide details of the three steps of the MH-within-Gibbs algorithm for posterior inference discussed in Section \ref{sec:inference} 
of the main paper. 
We focus on situations where the observations are multivariate binary outcomes. Generalization to outcomes with multiple categories is a straightforward adaptation.
Algorithm \ref{alg:inference} 
of the main paper gives a summary of the sampling steps.

\subsection{Metropolis-Hastings for Tree Topology and Divergence Times} \label{supp:inference:mh}

As described in Section 3.1.2 of \cite{yao2022probabilistic}, we first uniformly sample an internal node $u \in \{K+2, \ldots, 2K\}$, and split the current tree $\cT$ into two parts: a detached subtree $\cT_D$ rooted at the parent of the sampled node $u = pa(w)$, and the remaining tree $\cT_R$ after detaching $\cT_D$ from the current tree. Next, we simulate a new node $u'$ on $\cT_R$ at time $t_{u'}$ by following the branching process specified via divergence function $a(\cdot)$. A candidate tree $\cT'$ is formed by re-attaching the subtree $\cT_D$ to the remaining tree at node $u'$ and time $t_{u'}$. The re-attaching time should be no later than the detaching time $t_u$ (i.e. $t_{u'} < t_u$) to preserve branch lengths of $\cT_D$, except for the branch connected to its root $u$. The corresponding proposal distribution from $\cT$ to $\cT'$ is the probability of diverging at $u'$ on the tree $\cT_R$, and we denote the proposal distribution as $q(u', \cT_R)$. The MH acceptance probability is then
\begin{equation}
    \min \left\{1, \frac{ p \left( \cT', \bt' \mid \bOmega_{-(\cT',\bt')} \right) q(u, \cT_R) }{ p \left( \cT, \bt \mid \bOmega_{-(\cT,\bt)} \right) q(u', \cT_R) } \right\}. \label{eq:mh:prob}
\end{equation}

\begin{figure}[!h]
    \centering
    \includegraphics[width=0.8\linewidth]{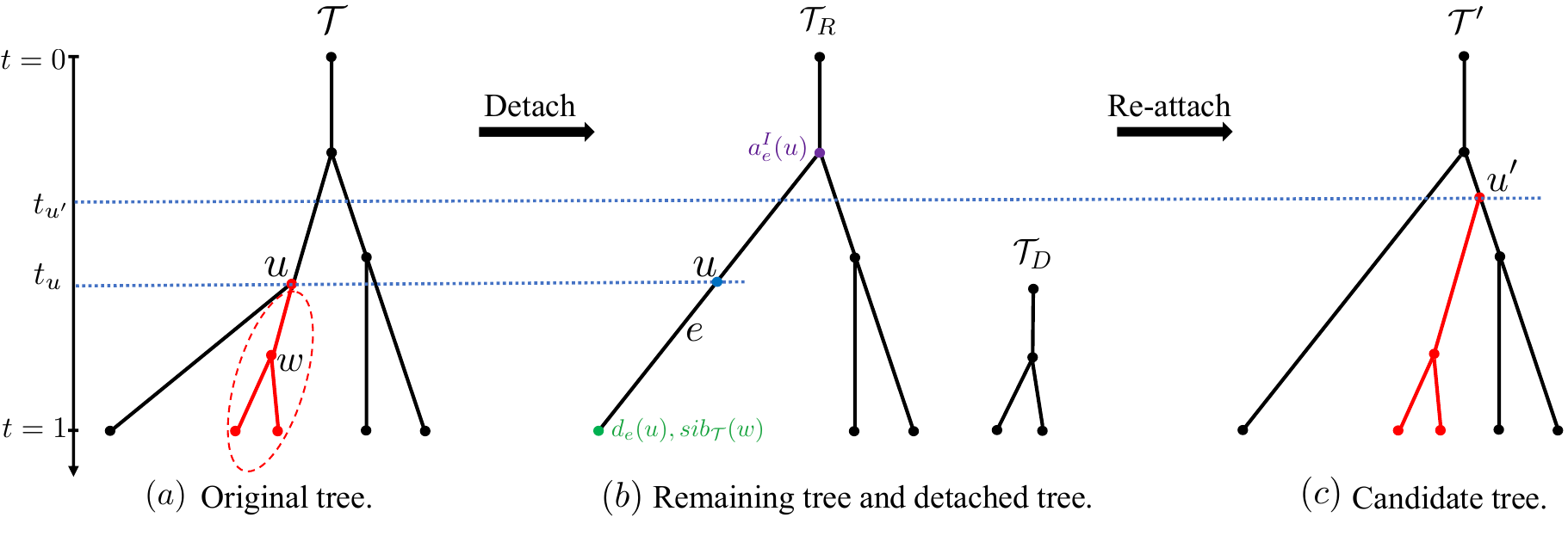}
    \caption{Example diagram of proposing a candidate tree $\cT'$ in the MH algorithm.}
    \label{fig:mh_tree}
\end{figure}

The target distribution is calculated as $p \left( \cT, \bt \mid \bOmega_{-(\cT,\bt)} \right) \propto p(\cT, \bt \mid c) p \left( \bm \eta \mid \cT, \bt, \bsigma^2 \right)$,
which is the same as \eqref{eq:ddt:jointdensity}.

To calculate the ratio between the proposal distributions of the current tree and the candidate tree, we are only concerned about the branch $e$ of $\cT_R$ from which the subtree $\cT_D$ is detached, because $\cT_R$ is shared by $\cT$ and $\cT'$. 
Denote $a_e^I(u) \subseteq {K+2, \ldots, 2K}$ as the internal nodes on the path from root to the detached point $u$ on $\cT_R$ earlier than $t_u$, and denote $d_e(u)$ as the nodes connected to $u$ on the branch $e$ of $\cT_R$ after $t_u$. 
On the current tree $\cT$, let the sibling node of $w$ be $sib(w)$, which is the child node of the detached point on $\cT_R$. Then the proposal distribution of the current tree is
\begin{align}
    q(u, \cT_R) \propto 
    \left[ \prod\limits_{r \in a_e^I(u)} \exp \left\{ [A(t_{pa(r)}) - A(t_{r})] / m(r) \right\} \right]
    \left[ \prod\limits_{r \in d_e(u)} \frac{ m(r) }{ m(pa(r)) } \right]
    \frac{ a (t_u) }{ m(sib(u)) }, \label{eq:mh:proposal}
\end{align}
where the $pa(r)$ is the parent node of $r$ on $\cT_R$, and $m(r)$ counts the number of leaves possessed by the subtree of $\cT_R$ rooted at $r$, or equivalently the number of ``particles" passing through $r$ on $\cT_R$. The first product term in equation \eqref{eq:mh:proposal} is the probability that no divergence happens on the path from root to $u$ on $\cT_r$, the second product term is the probability of selecting branches for particles to travel through in a self-reinforcing scheme, and the last term is the instantaneous probability of diverging at time $t_u$.
The proposal distribution of the candidate tree is obtained by replacing $\cT$ with $\cT'$ and $u$ with $u'$.

\subsection{Augmented Gibbs Sampler for \texorpdfstring{$\bm \eta$} {}}\label{supp:algorithm:eta}

For step (b), we sample the leaf parameter $\bm \eta$ by augmenting the conditional distribution $p \left( \bm \eta \mid \bOmega_{-\bm \eta}, \bY \right)$ in equation (7) of the main paper 
with two sets of auxiliary variables,
applying the approach in \cite{dalla2021polya}.
The first auxiliary variables that follow logistic distributions ease the implementation of the Gibbs sampler, in a similar manner to the probit regression model \citep{albert1993bayesian}.
The second auxiliary variables that follow P{\'o}lya-Gamma distributions provide an elegant closed-form solution to dealing with the logistic link \citep{polson2013bayesian}. We find empirically that incorporating both auxiliary variables significantly improves stability of the posterior chains, compared to the P{\'o}lya-Gamma technique alone. 

For individual $i \in [N]$ and item $j \in [J]$, let  
\begin{align}
    Y_{i,j} = I \{ W_{i,j} > 0 \}, \ \ 
    W_{i,j} \mid  Z_i = k \sim \mathrm{Logistic} \left( \eta_{j,k}, 1 \right), \label{eq:logistic_dist}
\end{align}
where $\mathrm{Logistic} \left(\mu, s \right)$ stands for logistic distribution with location parameter $\mu$ and scale parameter $s$, whose probability density function is $f(x; \mu, s) = \frac{\exp\{ (x-\mu) / s \}}{ s \left( 1 + \exp \{(x-\mu)/s \} \right)^2 }$, and cumulative distribution function is $H(x) = 1 / (1+\exp(-x))$ when $s = 1$.
We collect all elements $W_{i,j}$ into an $N \times D_0$ matrix $\bW$. The augmented posterior distribution is
\begin{align}
    p \left( \bm \eta, \bm W \mid \bOmega_{-\bm \eta}, \bY \right) &\propto 
    p(\bm \eta \mid \cT, \bt, \bsigma^2) p(\bY \mid \bW) P(\bW \mid \bZ, \bm \eta) \nonumber \\
    &= \cM \cN_{D_0 \times K} \left( \bm 0, \bOmega, \bSigma \right) \\
    &\ \ \cdot \prod\limits_{i=1}^N \prod\limits_{j=1}^{J} \left[ I \{ W_{i,j} > 0 \} Y_{i,j} + I \{ W_{i,j} \leq 0 \} (1 - Y_{i,j}) \right] \nonumber \\
    &\ \ \cdot \prod\limits_{i=1}^N \prod\limits_{k=1}^K 
    \left\{ \prod\limits_{j=1}^{J} \frac{ \exp ( w_{i,j} - \eta_{j,k}) }{ [1 + \exp ( w_{i,j} - \eta_{j,k})]^2 } \right\} ^{I\{Z_i = k\}}. \label{eq:posterior:eta:augmented}
\end{align}

We next deal with the logistic link by augmenting the distribution in equation \eqref{eq:posterior:eta:augmented} with P{\'o}lya-Gamma auxiliary variables, as proposed in \cite{polson2013bayesian}.
Specifically, we apply the following identity:
\begin{equation}
    \frac{(e^x)^a}{ (1 + e^x)^b } = 2^{-b} e^{\kappa x} \int_0^\infty \exp(- s x^2 / 2) p(s) ds, \label{eq:logitpglink}
\end{equation}
where $\kappa = a - b/2, a> 0, b > 0$, and $p(s)$ is the density function of a P{\'o}lya-Gamma (PG) distribution with shape parameter $b$ and exponential tilting parameter 0, denoted as PG$(b, 0)$. 

We apply the P{\'o}lya-Gamma identity in \eqref{eq:logitpglink} to the component in the curly bracket of the last line of \eqref{eq:posterior:eta:augmented}, which becomes
\begin{equation*}
    \prod\limits_{j=1}^{J} 2^{-2} \int_0^\infty \exp \left[- s_{i,j} ( w_{i,j} - \eta_{j,k})^2 / 2 \right] p(s_{i,j}) d s_{i,j},
\end{equation*}
where $s_{i,j} \sim \mathrm{PG} (2, 0)$. Note that $\exp \left(- s_{i,j} \eta_{j,k}^2 / 2 \right) p(s_{i,j})$ is the unnormalized density of a PG$(2, \eta_{j,k})$ random variable. Collecting all $s_{i,j}$ into an $N \times D_0$ matrix $\bS$, we obtain the augmented posterior distribution with the two sets of auxiliary variables as
\begin{align*}
    p \left( \bm \eta, \bm W, \bS \mid \bOmega_{-\bm \eta}, \bY \right) &\propto 
    p(\bm \eta \mid \cT, \bt, \bsigma^2) p(\bY \mid \bW) P(\bW, \bS \mid \bZ, \bm \eta) \nonumber \\
    &= \cM \cN_{D_0 \times K} \left( \bm 0, \bOmega, \bSigma \right) \nonumber \\
    &\ \ \cdot \prod\limits_{i=1}^N \prod\limits_{j=1}^{J} \left[ I \{ W_{i,j} > 0 \} Y_{i,j} + I \{ W_{i,j} \leq 0 \} (1 - Y_{i,j}) \right] \nonumber \\
    &\ \ \cdot \prod\limits_{i=1}^N \prod\limits_{k=1}^K \left\{ \prod\limits_{j=1}^{J} 2^{-2} \exp \left[- s_{i,j} ( w_{i,j} - \eta_{j,k})^2 / 2 \right] p(s_{i,j}) \right\} ^{I\{Z_i = k\}}.
\end{align*}

The full conditional distributions are derived as follows.

\textit{The conditional distribution of $\bW$:} The random variables $\{ W_{i,j} \}$ are mutually independent with conditional distributions
\begin{align*}
    & p \left( W_{i,j} \mid \bm \eta, \bZ, \bS, \bY \right) \\
    \propto & \exp \left\{ - \frac{1}{2 s_{i,j}^{-1} \left( w_{i,j} - \eta_{Z_i, j}\right)^2} \right\} \left[ I\{ w_{i,j} > 0\} Y_{i,j} + I\{ w_{i,j} \leq 0\} (1-Y_{i,j}) \right],
\end{align*}
which are truncated normal distributions. More precisely, the conditional distribution of $W_{i,j}$ is
\begin{align} \label{eq:cond:w}
    P \left( W_{i,j} \mid \bm \eta, \bZ, \bS, \bY \right) \sim 
    \begin{cases}
        \cN_{(0, \infty)} \left( \eta_{Z_i, j}, s_{i,j}^{-1} \right), &\text{ if } Y_{i,j} = 1 \\
        \cN_{(-\infty, 0]} \left( \eta_{Z_i, j}, s_{i,j}^{-1} \right), &\text{ if } Y_{i,j} = 0
    \end{cases},
\end{align}
where $\cN_{E} \left( \mu, \tau \right)$ indicates a normal distribution with mean $\mu$ and variance $\tau$ restricted to the interval $E$.

\textit{The conditional distribution of $\bS$:} The full conditional distributions of the P{\'o}lya-Gamma random variables $s_{i,j}$ are mutually independent and 
\begin{align} \label{eq:cond:s}
    P(s_{i,j} \mid \bm \eta, \bZ, \bS, \bY) &\propto \exp \left\{ -s_{i,j} \left( w_{i,j} - \eta_{Z_i, j}\right)^2 / 2 \right\} p(s_{i,j}) \nonumber \\
    &\sim \mathrm{PG} \left(2, w_{i,j} - \eta_{Z_i, j} \right).
\end{align}

\textit{The conditional distribution of $\bm \eta^L$:} 
The leaf parameters $\bm \eta$ of each item group $g$ are mutually independent. Let $\vvec{\bm \eta}$ denote the $K J_g \times 1$-vector created from stacking columns of $\bm \eta$, and let $\otimes$ denote the Kronecker product. The full conditional distribution of $\vvec{\bm \eta}$ is
\begin{align*}
    & P(\vvec{\bm \eta} \mid \bW, \bS, \bOmega_{-\bm \eta}, \bY) \\
    &\propto \exp \left\{ -\frac{1}{2} \vvec{\bm \eta}^\top \left( \bSigma  \otimes \sigma_g^2 \bI_{D_0} \right)^{-1} \vvec{\bm \eta} \right\}
    \prod\limits_{i=1}^N \prod\limits_{k=1}^K \left\{ \prod\limits_{j=1}^{J} \exp \left[- ( w_{i,j} - \eta_{j,k})^2 / 2 \right] \right\} ^{I\{Z_i = k\}} \\
    &= \exp \left\{ -\frac{1}{2} \vvec{\bm \eta}^\top \left( \bGamma_g + \bSigma  \otimes \sigma_g^2 \bI_{D_0} \right) \vvec{\bm \eta} + 
    \vvec{\bm \eta} \vvec{\bm \xi_g} \right\},
\end{align*}
where $\bxi_{g}$ denotes a $K \times J_g$ matrix whose $(k,j)$-th entry is $\sum\limits_{i=1}^N I\{Z_i = k\} u_{i,j} w_{i,j}$, and $\bGamma = \mathrm{diag} \left( \vvec{\bgamma} \right)$ with $\bgamma_g$ being a $K \times J_g$ matrix whose $(k,j)$-th entry is $\sum\limits_{i=1}^N I\{Z_i = k\} u_{i,j}$.
Therefore, the conditional distribution of $\bm \eta$ is
\begin{align} \label{eq:cond:etaL}
    P(\bm \eta \mid \bW, \bU, \bOmega_{-\bm \eta}, \bY) \sim \cN_{KD_0} \left( \bmu, \bPsi \right),
\end{align}
where
\begin{align*}
    \bPsi_g = \left( \bD_g +  \bI_{J_g} \otimes \sigma_g^{-2} \bSigma_g^{-1} \right)^{-1}, \ \ 
    \bmu_g = \bPsi_g \vvec{\bm \xi_g^{(1)}}.
\end{align*}

\subsection{Gibbs Sampler for the Remaining Parameters} \label{supp:inference:otherparams}
Apart from the model parameters discussed in the previous sections, we derive the divergence hyperparameter $c$ and diffusion variance $\bsigma^2$ of the DDT process, as well as the latent class indicators $\bZ$ and class prevalence $\bpi$ in this section. Utilizing equation \eqref{eq:ddt:jointdensity}, the full conditional distribution of the divergence hyperparameter $c$ is
\begin{align}
    p(c \mid \bOmega_{-c}) &\propto p(\cT, \bt \mid c) \ p(c) \ \propto \prod\limits_{[uv] \in \cS(\cT)} c (1 - t_v)^{ c J_{v} - 1} c^{\alpha_c - 1} e^{- c \beta_c} \nonumber \\
    &= c^{K-1 + \alpha_c - 1} \exp \left\{ -c \left( \beta_c - \sum_{u\in \cV^I} J_u \log (1-t_u) \right) \right\} \nonumber \\
    & \sim G\left(K-1 + \alpha_c, \beta_c - \sum_{u\in \cV^I} J_u \log (1-t_u) \right). \label{eq:cond:c}
\end{align}

The full conditional distributions of $\bsigma_g^2$ of each item group $g$ are mutually independent with
\begin{align}
    p(\sigma_g^2 \mid \bOmega_{-c}) &\propto p(\bm \eta \mid \cT, \bt, \sigma_g^2) p(\sigma_g^2) \nonumber \\
    &\propto \left( \sigma_g^2 \right)^{- J_g K / 2} \exp \left\{ -\frac{1}{2} \sigma_g^{-2} \mathrm{Tr} \left((\bm \eta)^\top \bSigma^{-1} (\bm \eta) \right) \right\} \left( \sigma_g^2 \right)^{- \alpha_{\sigma_g} - 1} \exp \left(- \beta_{\sigma_g} / \sigma_g^2 \right) \nonumber \\
    &= \left( \sigma_g^2 \right)^{- J_g K / 2 - \alpha_{\sigma_g} - 1}
    \exp \left(- (\beta_{\sigma_g} + S_g / 2 ) \sigma_g^{-2} \right) \nonumber \\
    &\sim IG \left( J_g K / 2 + \alpha_{\sigma_g}, \beta_{\sigma_g} + S_g / 2 \right), \label{eq:cond:sigma}
\end{align}
where $S_g = \mathrm{Tr} \left((\bm \eta)^\top \bSigma^{-1} (\bm \eta) \right)$ and $\mathrm{Tr}(\bX)$ denotes the trace of a square matrix $\bX$. 

The full conditional distribution of $\bZ_i$ can be obtained from
\begin{align*}
    p(Z_i = k \mid \bOmega_{-\bZ}, \bY) &\propto \pi_k \prod\limits_{j=1}^{J} (\theta_{j,k})^{Y_{i,j}} (1 - \theta_{j,k})^{1 - Y_{i,j}},
\end{align*}
which is a categorical distribution on $[K]$ with the probability of taking value $k$ being 
\begin{equation} \label{eq:cond:z}
\frac{ \pi_k \prod\limits_{j=1}^{J} (\theta_{j,k})^{Y_{i,j}} (1 - \theta_{j,k})^{1 - Y_{i,j}} }{ \sum\limits_{l = 1}^K \pi_l \prod\limits_{j=1}^{J} (\theta_{l,j})^{Y_{i,j}} (1 - \theta_{l,j})^{1 - Y_{i,j}}}.
\end{equation}

\section{Additional Simulation Study Details} \label{supp:sec:simulation}
The same hyperparameter setting in the priors of $\sigma_g^2$, $c$, and $\bpi$ applies to the synthetic (Section \ref{sec:sim:synthetic} of the main paper) and semi-synthetic (Section \ref{sec:sim:real} of the main paper) data simulations and real data application (Section \ref{sec:data} of the main paper). Specifically, the priors for all $\sigma_g^2$ are specified as Inverse-Gamma$(2, 2)$, the prior for $c$ is Gamma$(1, 1)$ and the prior for $\bpi$ is Dirichlet$(5, \ldots, 5)$. For all simulation scenarios, we run the sampling algorithm for 8,000 iterations and discard the first 5,000 burn-in samples for each simulated dataset. Label switching is addressed post hoc via the Equivalence Classes Representative \citep[ECR,][]{papastamoulis2014HandlingLabel}.

\subsection{Simulation I: Synthetic Data Setup and Results} \label{supp:sec:simulation:syntheticsetup}

\begin{figure}[h!]
    \centering
    \includegraphics[width=0.8\linewidth]{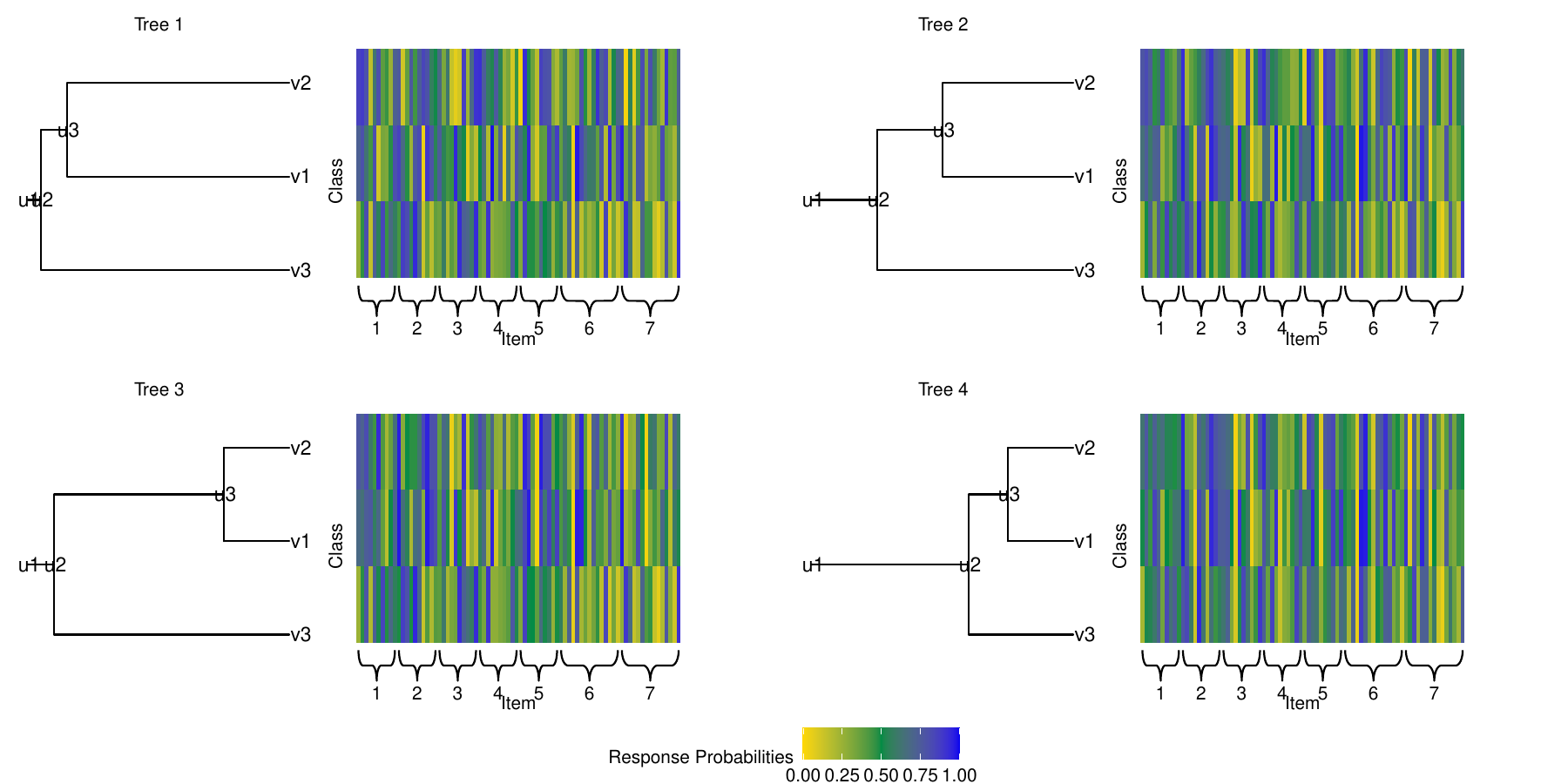}
    \vspace{-3mm}
    \caption{Simulation I: Four tree structures, ordered by increasing between-class correlation, in the synthetic data simulation. For each tree, the heatmap on the right display one of the 100 simulated sets of item response probabilities $\bTheta$. Beneath the heatmap are major food group indices of the $J = 80$ items in the columns.}
    \label{fig:sim:true_trees}
\end{figure}

Four tree structures are considered, as shown in Figure \ref{fig:sim:true_trees}.
The class prevalence is set to $\bpi = (0.4, 0.3, 0.3)$, and the group-specific diffusion variances are $\sigma_g^2 = 0.6^2$ for $g \leq 5$ and $\sigma_g^2 = 2^2$ for $g = 6, 7$. For each tree, we simulate 20 sets of response probabilities, for each of which we simulate 5 independent datasets of multivariate binary responses for $N \in \{100, 200, 400\}$ individuals following LCMs; hence, a total of 100 independent datasets for each $N$. 
For the ``DDT-LCM (misspecified tree)" method, the misspecified tree is fixed at the structure Tree 4 when we perform posterior sampling of the Tree 1 and Tree 2 scenarios, and the misspecified tree for Tree 3 and Tree 4 scenarios is fixed at the structure of Tree 1.

\begin{figure}[h!]
    \centering
    \includegraphics[width=\linewidth]{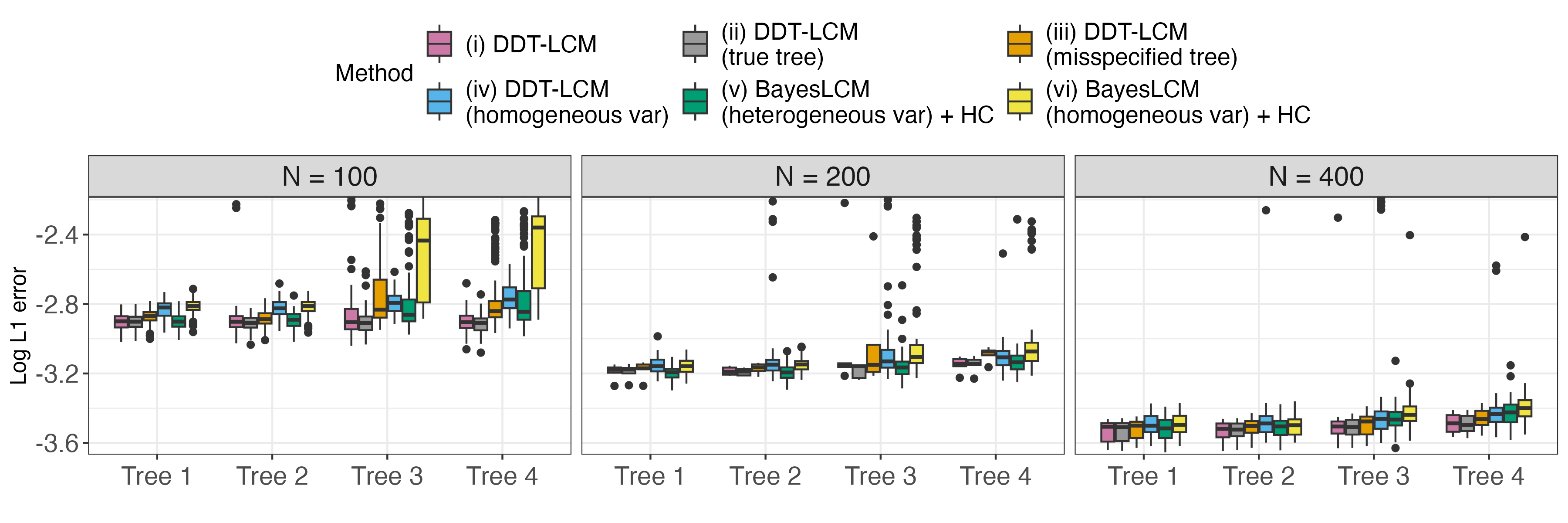}
    \vspace{-3mm}
    \caption{Simulation I: Average log absolute biases of $\bTheta$ calculated as $\mathrm{Bias} \left( \hat{\bTheta} \right) = (KJ)^{-1} \sum\limits_{k=1}^K \sum\limits_{g = 1}^G \sum\limits_{j = 1}^{J_g} \left| \hat{\theta}_{g,j,k} - \theta_{g,j,k} \right|$.}
    \label{fig:sim:synthetic_bias}
\end{figure}

Bayesian shrinkage allows for borrowing information across latent classes, although sometimes at the price of introducing biases.
Here we assess the biases in estimating the conditional probabilities $\bTheta$. 
Figure \ref{fig:sim:synthetic_bias} suggests that the relative model performances in terms of absolute biases are the same as those in terms of RMSE in the second row of Figure \ref{fig:sim_boxplot_combined} in the main paper.
The lower biases of (i) than (iv) imply that jointly learning between-class similarity and the guiding tree improves parameter estimation. A known true tree structure reduces biases in estimating $\bTheta$, when comparing (i) and (ii). If the tree is misspecified at a structure far from the truth, biases may be larger than DDT-LCM (i) and no smaller than the plain BayesLCM (v).
Additionally, DDT-LCM has smaller biases than other baseline models for weaker separation between the classes (from Tree 1 to 4). Larger sample sizes tend to make up for the performance disadvantage of other methods ($N=100$ to $400$).

\begin{figure}[!ht]
    \centering
    \begin{minipage}[c]{\textwidth}
        \centering
        \includegraphics[width=0.9\textwidth]{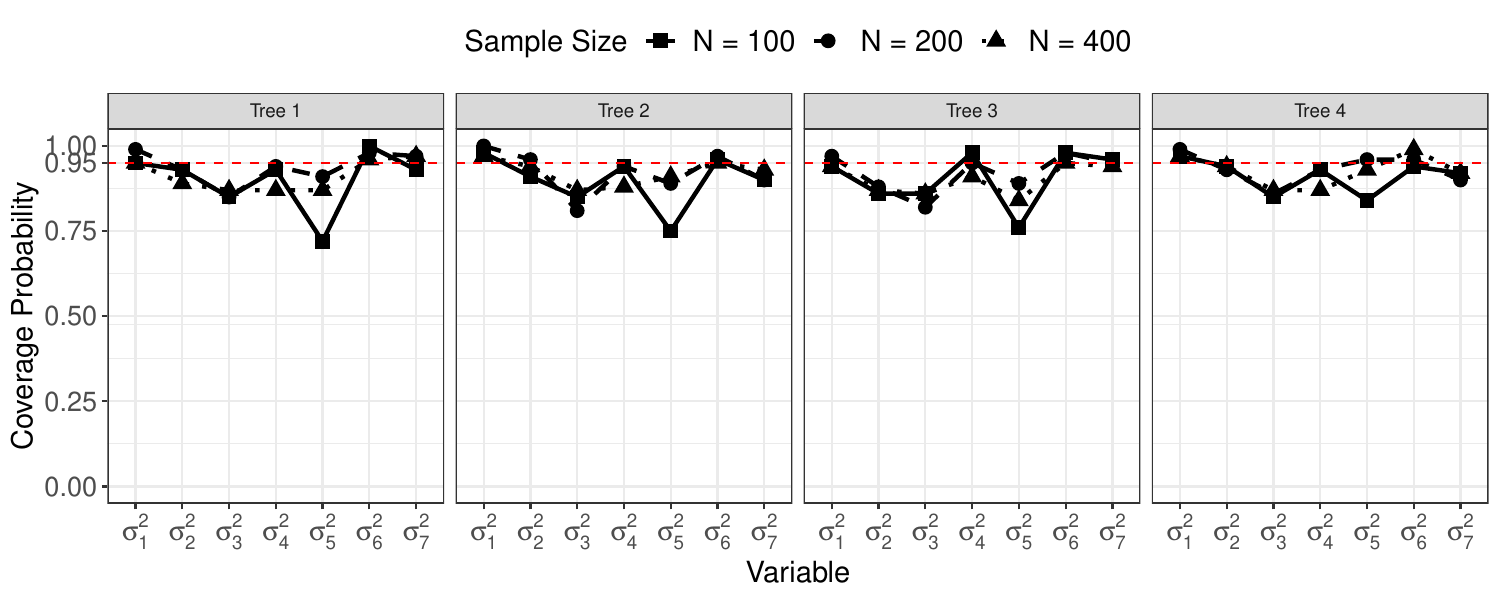}
        \label{fig:sim_K3_sigma_coverage}
    \end{minipage}
    \begin{minipage}[c]{\textwidth}
        {\small (a) Empirical coverage probabilities of the approximate 95\% credible intervals The horizontal dashed lines indicate 95\%.}
    \end{minipage}
    \begin{minipage}[c]{\textwidth}
        \centering
        \includegraphics[width=0.8\textwidth]{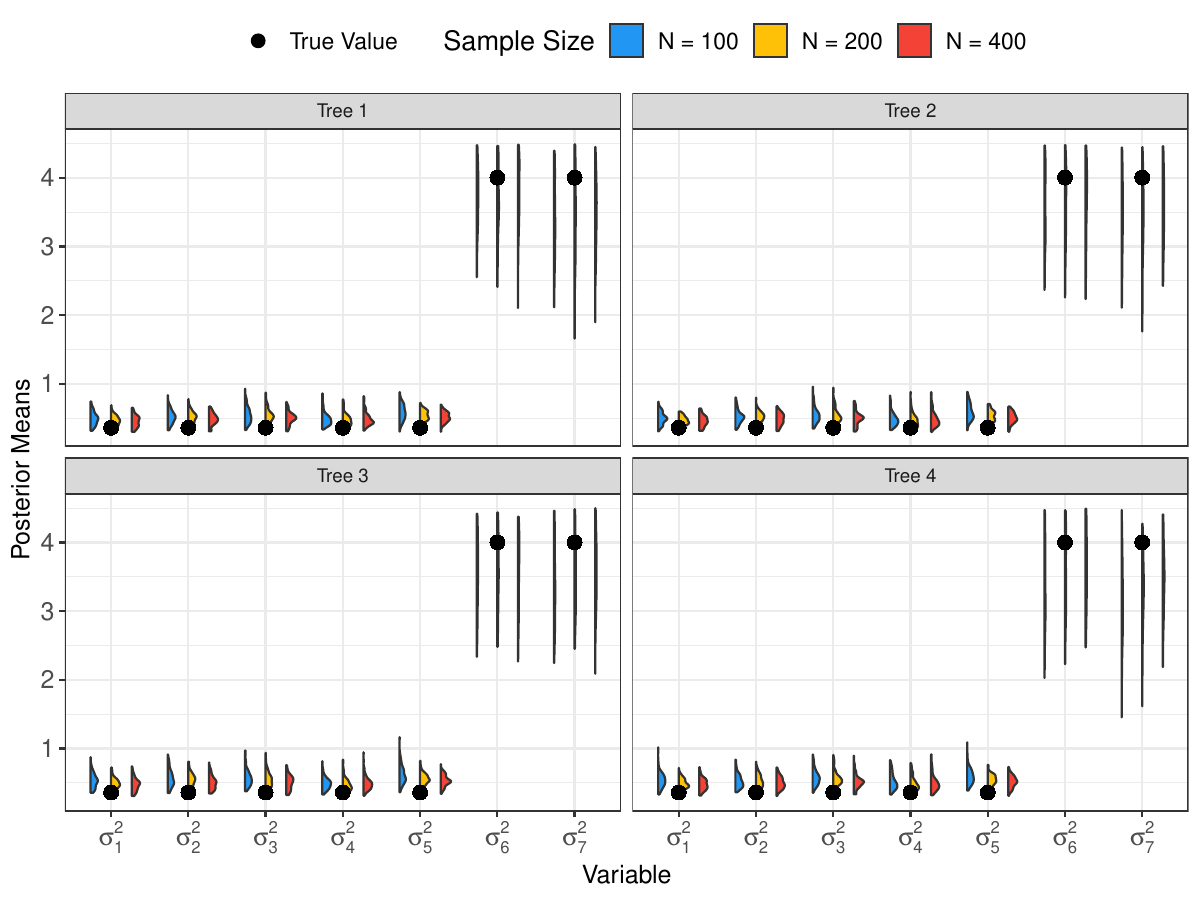}
        \label{fig:sim_K3_sigma_distribution}
    \end{minipage}
    \begin{minipage}[c]{\textwidth}
        {\small (b) Distribution of posterior mean estimates of $\bsigma^2$. For each $\sigma_g^2$, the three distributions correspond to sample sizes $N = 100, 200$, and 400 from left to right.}
    \end{minipage}
    \caption{Simulation I: results of estimating group-specific diffusion variance parameter $\bsigma^2$, based on DDT-LCM from 100 datasets under different tree and sample size scenarios.}
    \label{fig:sim_K3_sigma}
\end{figure}

\change{
\paragraph*{Inference Results of Diffusion Variances}

Figure \ref{fig:sim_K3_sigma} shows the coverage probabilities and the distributions of posterior means of the diffusion variances $\bsigma^2$ across 100 replications. We see that for all the seven major food groups, the posterior means of $\sigma_g^2$ are close to the true parameter values.
}

\subsection{Simulation II: Semi-Synthetic Data} \label{sec:sim:real}


\subsubsection{Simulation Setup} \label{sec:sim:real:setup}

We set the true tree as the MCC tree (left of Figure \ref{fig:data_result} of the main paper 
) estimated by DDT-LCM from the HCHS/SOL data. The MCC tree implies weakly separated classes with between-class correlations at least 0.8. 
We set the class prevalence to be the posterior mean  estimated from the HCHS/SOL data analysis, which is $\bpi = (0.14, 0.14, 0.2, 0.18, 0.17, 0.16)$. The group-specific diffusion variances are $\sigma_g^2 = 1^2$ for $g \leq 5$ and $\sigma_g^2 = 2.3^2$ for $g = 6, 7$. Based on the MCC tree (Figure 4 of the main paper 
), we simulate 20 sets of response probabilities, for each of which we simulate 5 independent datasets of multivariate binary responses for $N \in \{400, 800\}$ individuals following LCMs; hence, a total of 100 independent datasets for each $N$. 
Figure \ref{fig:K6_misspecified_tree} displays the misspecified tree structured used in method ``DDT-LCM (misspecified tree)".

\begin{figure}[h!]
    \centering
    \includegraphics[width=0.8\linewidth]{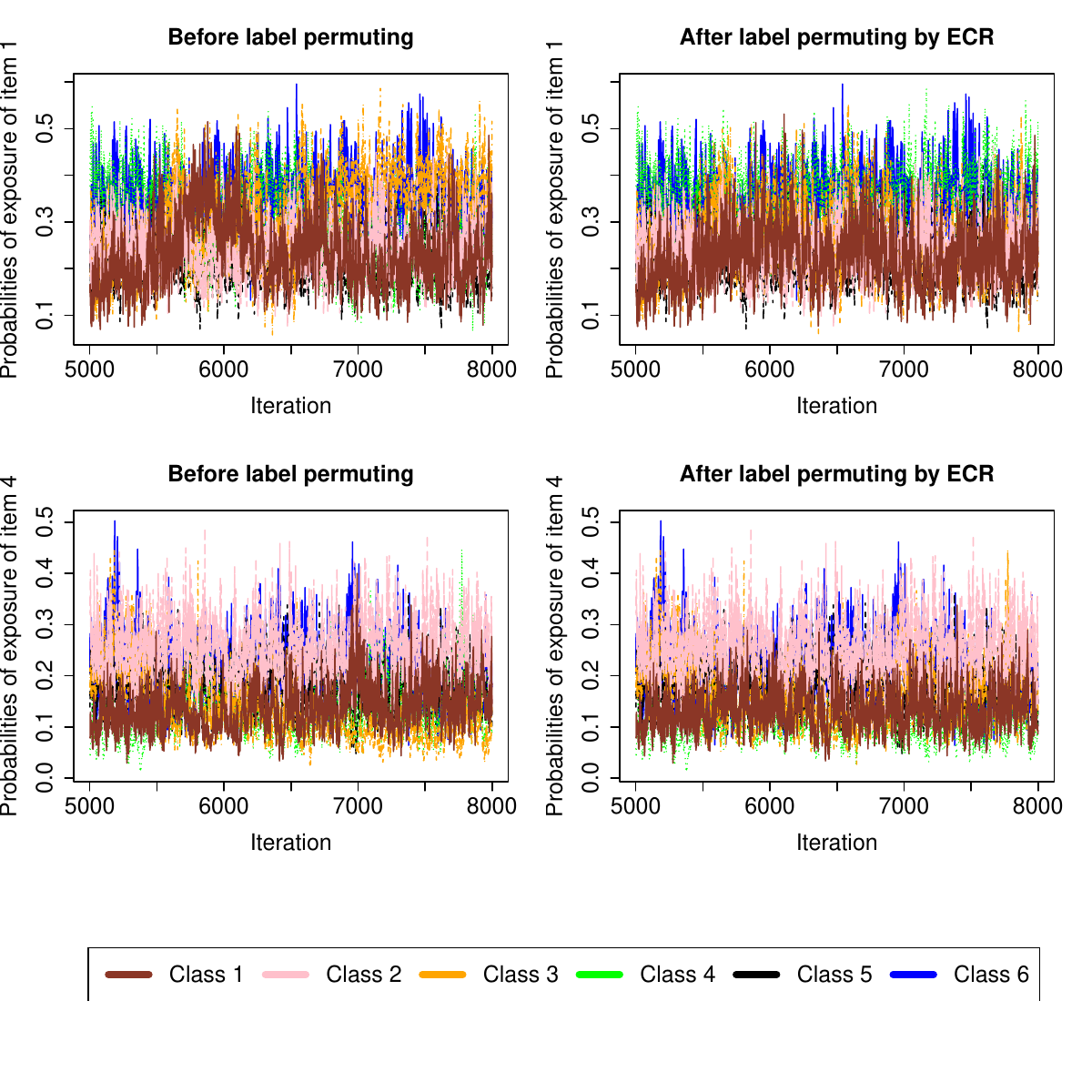}
    \vspace{-3mm}
    \caption{Simulation II: reordered MCMC output of the probabilities of exposure to items 1 $(\theta_{k,1,1})$ and 4 $(\theta_{k,1,4})$ in one replication of DDT-LCM, $k = 1, \ldots, 6$. The trace plots display the 3000 posterior samples after 5000 burn-in's.}
    \label{fig:sim:label_switching}
\end{figure}

The performance of correcting label switching by the ECR algorithm can be examined by the trace plots of the probabilities of exposure to items, as displayed in Figure \ref{fig:sim:label_switching}. We see that the posterior chains mix well after label correction. 

\change{
The computation time for drawing 8000 posterior samples with the semi-synthetic simulation setup cost 38 minutes for the proposed DDT-LCM and 5.3 minutes for the classical Bayesian LCM, on average across 100 replications. The computation was conducted in R version 4.2.0 on a 2018 MacBook Pro.
}

\subsubsection{Data-driven choice of $K$} \label{sec:sim:chooseK}
The performance of DDT-LCM in the simulations has been demonstrated under a known $K$. To provide a practical estimation pipeline applicable to real-world data, we can select $K$ in a data-driven manner. Our rationale here is to apply a practically useful criterion that leans towards a model with good out-of-sample predictive performance while remaining parsimonious. 
Viewing $K$ as a hyperparameter, we choose the $K$ that yields the average largest predictive log-likelihood on validation datasets from 5-fold cross-validation. 
Specifically, the observations of $N$ individuals are randomly split into a training set and a testing set according to a 4:1 ratio. For the $s$-th training set, $s \in [5]$, we apply the Gibbs sampler and obtain the posterior means of the class prevalences $\hat{\pi}$ and response probabilities $\hat{\theta}$. The predictive likelihood on the corresponding testing set is computed as 
\begin{equation} \label{eq:crossvalidation}
    p_{K,s}^{test} = \prod\limits_{i \in I^{test}_s} \left[ \sum\limits_{k=1}^K \hat{\pi}_k \prod\limits_{j=1}^{J} \left( \hat{\theta}_{j,k} \right)^{Y_{i,j}} \left( 1 - \hat{\theta}_{j,k} \right)^{1 - Y_{i,j}} \right],
\end{equation}
where $I^{test}_s$ denotes the indices of individuals belonging to the $s$-th testing set. The average predictive log-likelihood is then calculated as $l_{K}^{test} = \frac{1}{5} \sum\limits_{s=1}^5 \log p_{K,s}^{test}$. The model with a larger $l_{K}^{test}$ is preferred. 

We propose to use the cross-validated predictive loglikelihood (CV-PL) in equation \ref{eq:crossvalidation} as our main criterion. Specifically, we perform 5-fold CV to achieve a computationally feasible criterion, as opposed to leave-one-out cross-validation (LOO-CV).
We conduct a simulation study to evaluate the performance of $l_{K}^{test}$ on the selection of $K$, focusing on semi-synthetic scenario with the truth $K_{\mathrm{true}} = 6$. For each candidate number of classes $K_{\mathrm{candidate}}$ from $\{4,5,6,7,8\}$, we compute the above measure $l_{K_{\mathrm{candidate}}}^{test}$. We find that for $N = 400$, the percentages of datasets that each of $K = 4, 5, 6, 7, 8$ is selected are 11\%, 36\%, 45\% (truth), 7\%, and 1\%, respectively; for $N = 800$, the percentages are 2\%, 15\%, 73\% (truth), 8\%, and 2\%, respectively. This result indicates that the predictive log-likelihood is a reasonable metric to select the best $K$, and the selection accuracy increases with sample size. While we acknowledge that selecting the correct $K$ is not an easy task under the relatively small sample size $N = 496$ in our real data application, the practical utility of our procedure is still valuable in analysis of real-world datasets. Developing a rigorous understanding of the theoretical behavior of this measure requires further investigation.

\change{
\paragraph{Other Criteria for Choosing $K$} \label{sec:sim:chooseK:other}
In our experiments, we have also compared several other model selection criteria in Table \ref{tab::freq_table}, including log pointwise predictive density (LPPD), AIC \citep{wagenmakers2004aic}, 
Deviance Information Criterion (DIC, \cite{spiegelhalter2002bayesian}), and Widely Applicable Information Criterion (WAIC, \cite{watanabe2010asymptotic}). See \cite{gelman2014understanding} for a comprehensive overview of these commonly applied predictive information criteria for evaluating Bayesian models. Here we present a brief comparison bewteen these criteria in the context of DDT-LCM. All these criteria are design to approximate the expected LPPD.
\begin{itemize}
\item \textbf{CV-PL} captures out-of-sample prediction error and avoids overfitting with a large $K$. LOO-CV may provide a stable estimate but requires $N$ data partitions and is computationally intensive. A smaller-fold CV (such as five-fold in our criterion) balances computational requirements and predictive performance.

\item \textbf{LPPD} measures the predictive performance for a new data produced from the true data-generating process. The LPPD is computed based on draws from the posterior distribution. The LPPD of observed data overestimates the expected LPPD for future data. AIC, DIC, and WAIC are designed to correct the bias of LPPD when approximating the expected LPPD.

\item \textbf{AIC} performs bias correction based on adymptotic normal posterior distribution by subtracting the number of parameters from the log predictive density, conditional on the maximum likelihood estimate (MLE). AIC is asymptotically equivalent to LOO-CV computed using the MLE.

\item \textbf{DIC} is a Bayesian version of AIC by replacing the MLE with the posterior mean and replacing the number of parameters with a data-based bias correction. Similar to AIC, bias correction is achieved based on the asymptotic posterior $\chi^2$ distribution. DIC is asymptotically equivalent to LOO-CV computed using predictive densities.

\item \textbf{WAIC} is a fully Bayesian criterion that corrects the effective number of parameters to address overfitting. Instead of conditioning on a point estimate like AIC and DIC, WAIC averages over the posterior distribution. WAIC is asymptotically equal to LOO-CV.
\end{itemize}

In Table \ref{tab::freq_table}, ELPPD underestimates $K$ when the sample sizes are $N = 400$ and 800. WAIC1 and DIC1 are computed using average posterior loglikelihood, WAIC2 and DIC2 are computed using variance of posterior loglikelihood (See \cite{gelman2014understanding} for details). The two computation methods do not produce a big difference in results in our settings. The two WAICs tend to underestimate $K$, but with a slightly increased tendency to choose a large $K = 8$ compared to ELPPD or DIC. The two DICs have a higher chance to correctly select $K = 6$ when $N = 800$ compared to $N = 400$. However, none of these alternative criteria perform as well as CV-PL. This is likely because the asymptotic criteria WAIC and DIC may not be effective under the small sample size settings with weakly separated latent classes considered in our simulations.

\begin{table}[H]
    \centering
    \caption{Frequency of choosing $K$ using different criteria} \label{tab::freq_table}
    \begin{tabular}{cccccccc}
      \toprule
    $N$ & $K$ & \textbf{CV-PL} & ELPPD & WAIC1 & WAIC2 & DIC1 & DIC2 \\ 
      \midrule
    \multirow{5}{*}{400} & 4 &  11 &  79 &  41 &  56 &   40 &  42 \\ 
       & 5 &  36 &  14 &  17 &   9 &   39 &  35 \\ 
       & 6 &  45 &   5 &  11 &   4 &   15 &  17 \\ 
       & 7 &   7 &   2 &  16 &  14 &  3 &   3 \\ 
       & 8 &   1 &   0 &  15 &  17 &  3 &   3 \\ 
       \midrule
    \multirow{5}{*}{800} & 4 &   2 &  72 &  47 &  60 &   32 &  29 \\ 
       & 5 &  15 &  13 &  19 &  17 &   29 &  33 \\ 
       & 6 &  73 &  10 &   6 &   5 &   27 &  23 \\ 
       & 7 &   8 &   4 &   9 &   2 &  5 &  10 \\ 
       & 8 &   2 &   1 &  19 &  16 &  7 &   5 \\ 
       \bottomrule
    \end{tabular}
\end{table}

}

\subsubsection{Choice of diffusion variance} \label{sec:simulation:real:variance}
\begin{figure}[H]
\centering
\begin{minipage}{.4\textwidth}
    \centering
    \includegraphics[width=\linewidth]{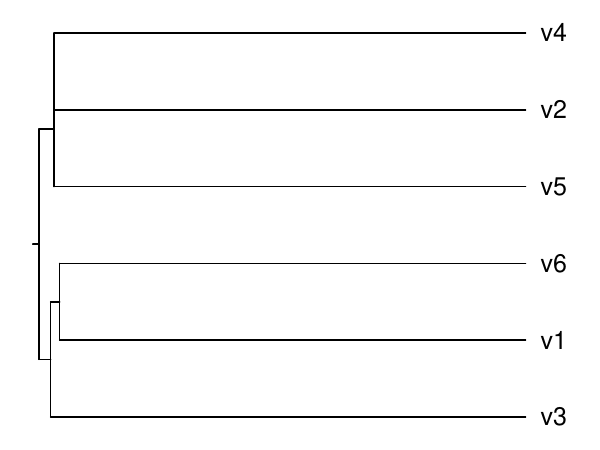}
\end{minipage}\qquad \qquad
\begin{minipage}{.4\textwidth}
    \centering
    \includegraphics[width=\linewidth]{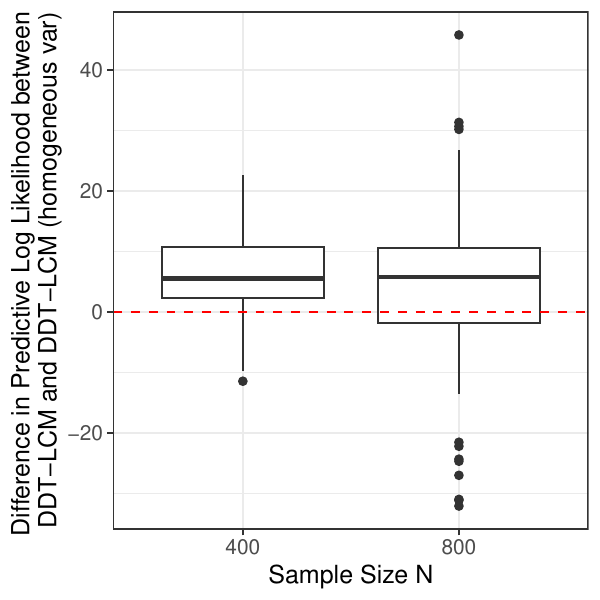}
\end{minipage}

\bigskip

\begin{minipage}[H]{.45\textwidth}
\centering
\caption{Simulation II: Misspecified tree structure for method ``DDT-LCM (misspecific tree)".}
\label{fig:K6_misspecified_tree}
\end{minipage}\qquad
\begin{minipage}[H]{.45\textwidth}
\centering
\caption{Simulation II: log-likelihood comparison to choose group-specific or homogeneous diffusion variance parameters.  The red dashed line indicates 0.}
\label{fig:sigma_llk_difference}
\end{minipage}
\end{figure}

We provide a practical procedure to select group-specific or diffusion variance specification when analyzing real data. For DDT-LCM and ``DDT-LCM (homogeneous var)" methods under a pre-specified $K$, we choose the method that produces a higher average predictive log-likelihood from cross-validation. Figure \ref{fig:sigma_llk_difference} shows the comparison results of the two methods over 100 simulated semi-synthetic datasets. For both sample sizes $N = 400$ and 800, DDT-LCM with group-specific diffusion variance parameters performs better than the one with homogeneous variance parameters.

\newpage
\section{Additional Results in HCHS/SOL Data Analysis} \label{supp:item_list}
		
\begingroup
\scriptsize
\begin{longtable}{|l|p{0.09\textwidth}|p{0.47\textwidth}|p{0.05\textwidth}|p{0.05\textwidth}|p{0.08\textwidth}|}
\caption{Food items included in data application to HCHS/SOL 24-hour dietary recall data. 
Daily consumed foods: (High) at least one serving a day, and (Low) less than one serving a day. 
All other foods: (+) any consumption, and (-) no consumption. 
The last column indicates the percentages of individuals having positive exposure to the food items. 
Only foods with $\geq 5\%$ or $\leq 95\%$ level 1 exposure were included in the analysis. }  \label{tab:item_labels} \\
  \hline 
  Group & Item Label & Description & Level 0 & Level 1 & \% Level 1 \\
  \hline\hline 
  \endfirsthead
\multicolumn{6}{c}%
{\tablename\ \thetable\ -- \textit{Continued from previous page}} \\
  \hline 
  Group & Item Label & Description & Level 0 & Level 1 & \% Level 1 \\
  \hline\hline
  \endhead
  \hline
  \endfoot 
  \hline
  \endlastfoot 
  \hline\multirow{11}{*}{Dairy} & diary\_1 & Milk - Whole & - & + & 49.19 \\ 
   & diary\_2 & Milk - Low Fat and Fat Free & - & + & 44.15 \\ 
   & diary\_3 & Ready-to-drink Flavored Milk - Reduced Fat & - & + & 14.31 \\ 
   & diary\_4 & Ready-to-drink Flavored Milk - Low Fat and Fat Free & - & + & 6.05 \\ 
   & diary\_5 & Sweetened Flavored Milk Beverage Powder with Non-fat Dry Milk & - & + & 63.10 \\ 
   & diary\_6 & Artificially Sweetened Flavored Milk Beverage Powder with Non-fat Dry Milk & - & + & 30.24 \\ 
   & diary\_7 & Yogurt - Sweetened Low Fat & - & + & 6.05 \\ 
   & diary\_8 & Yogurt - Sweetened Fat Free & - & + & 9.68 \\ 
   & diary\_9 & Yogurt - Nondairy & - & + & 5.85 \\ 
   & diary\_10 & Dairy-based Artificially Sweetened Meal Replacement/Supplement & - & + & 10.89 \\ 
   & diary\_11 & Infant Formula & - & + & 33.67 \\ 
  \hline\multirow{7}{*}{Fat} & fat\_1 & Cream - Reduced Fat & - & + & 11.90 \\ 
   & fat\_2 & Margarine - Regular & - & + & 13.10 \\ 
   & fat\_3 & Butter and Other Animal Fats - Reduced Fat & - & + & 24.40 \\ 
   & fat\_4 & Salad Dressing - Regular & - & + & 32.86 \\ 
   & fat\_5 & Salad Dressing - Reduced Fat/Reduced Calorie/Fat Free & - & + & 6.05 \\ 
   & fat\_6 & Gravy - Regular & - & + & 17.34 \\ 
   & fat\_7 & Gravy - Reduced Fat/Fat Free & - & + & 12.70 \\ 
  \hline\multirow{7}{*}{Fruit} & fruit\_1 & Citrus Juice & - & + & 13.10 \\ 
   & fruit\_2 & Fruit Juice excluding Citrus Juice & - & + & 57.46 \\ 
   & fruit\_3 & Citrus Fruit & - & + & 29.64 \\ 
   & fruit\_4 & Fruit excluding Citrus Fruit & - & + & 47.38 \\ 
   & fruit\_5 & Avocado and Similar & - & + & 57.86 \\ 
   & fruit\_6 & Fried Fruits & - & + & 41.94 \\ 
   & fruit\_7 & Fruit-based Savory Snack & - & + & 47.38 \\ 
  \hline\multirow{16}{*}{Grain} & grain\_1 & Grains, Flour and Dry Mixes - Whole Grain & - & + & 11.49 \\ 
   & grain\_2 & Grains, Flour and Dry Mixes - Some Whole Grain & - & + & 20.36 \\ 
   & grain\_3 & Grains, Flour and Dry Mixes - Refined Grain & Low & High & 15.73 \\ 
   & grain\_4 & Loaf-type Bread and Plain Rolls - Refined Grain & - & + & 12.50 \\ 
   & grain\_5 & Other Breads (quick breads, corn muffins, tortillas) - Some Whole Grain & - & + & 63.91 \\ 
   & grain\_6 & Other Breads (quick breads, corn muffins, tortillas) - Refined Grain & - & + & 20.56 \\ 
   & grain\_7 & Crackers - Whole Grain & - & + & 8.47 \\ 
   & grain\_8 & Crackers - Some Whole Grain & - & + & 30.04 \\ 
   & grain\_9 & Pasta - Whole Grain & - & + & 5.44 \\ 
   & grain\_10 & Ready-to-eat Cereal (not presweetened) - Whole Grain & - & + & 6.85 \\ 
   & grain\_11 & Ready-to-eat Cereal (presweetened) - Whole Grain & - & + & 34.27 \\ 
   & grain\_12 & Ready-to-eat Cereal (presweetened) - Some Whole Grain & - & + & 66.33 \\ 
   & grain\_13 & Cakes, Cookies, Pies, Pastries, Danish, Doughnuts and Cobblers - Some Whole Grain & - & + & 46.17 \\ 
   & grain\_14 & Snack Bars - Whole Grain & - & + & 10.89 \\ 
   & grain\_15 & Snack Chips - Whole Grain & - & + & 33.67 \\ 
   & grain\_16 & Flavored Popcorn & - & + & 15.32 \\ 
  \hline\multirow{12}{*}{Meat} & meat\_1 & Lamb & - & + & 44.76 \\ 
   & meat\_2 & Lean Lamb & - & + & 15.12 \\ 
   & meat\_3 & Game & - & + & 15.12 \\ 
   & meat\_4 & Poultry & - & + & 33.06 \\ 
   & meat\_5 & Lean Poultry & Low & High & 20.36 \\ 
   & meat\_6 & Fried Chicken - Commercial Entr\'ee and Fast Food & - & + & 14.72 \\ 
   & meat\_7 & Lean Fish - Fresh and Smoked & - & + & 7.06 \\ 
   & meat\_8 & Fried Fish - Commercial Entr\'ee and Fast Food & - & + & 20.16 \\ 
   & meat\_9 & Fried Shellfish - Commercial Entr\'ee and Fast Food & - & + & 11.69 \\ 
   & meat\_10 & Lean Cold Cuts and Sausage & - & + & 9.27 \\ 
   & meat\_11 & Nuts and Seeds & - & + & 30.04 \\ 
   & meat\_12 & Nut and Seed Butters & - & + & 23.19 \\ 
  \hline \multirow{15}{*}{Sugar} & sugar\_1 & Sugar & Low & High & 36.49 \\ 
   & sugar\_2 & Syrup, Honey, Jam, Jelly, Preserves & - & + & 46.77 \\ 
   & sugar\_3 & Sauces, Sweet - Reduced Fat/Reduced Calorie/Fat Free & - & + & 11.49 \\ 
   & sugar\_4 & Chocolate Candy & - & + & 23.79 \\ 
   & sugar\_5 & Sweetened Soft Drinks & - & + & 6.25 \\ 
   & sugar\_6 & Artificially Sweetened Soft Drinks & - & + & 12.50 \\ 
   & sugar\_7 & Sweetened Fruit Drinks & - & + & 50.20 \\ 
   & sugar\_8 & Artificially Sweetened Fruit Drinks & - & + & 6.45 \\ 
   & sugar\_9 & Artificially Sweetened Tea & - & + & 51.21 \\ 
   & sugar\_10 & Sweetened Coffee & - & + & 7.46 \\ 
   & sugar\_11 & Unsweetened Coffee & - & + & 9.27 \\ 
   & sugar\_12 & Sweetened Coffee Substitutes & - & + & 36.69 \\ 
   & sugar\_13 & Unsweetened Coffee Substitutes & - & + & 50.00 \\ 
   & sugar\_14 & Nondairy-based Unsweetened Meal Replacement/Supplement & - & + & 13.71 \\ 
   & sugar\_15 & Miscellaneous Dessert & - & + & 11.69 \\ 
  \hline\multirow{10}{*}{Vegetable} & veg\_1 & Dark-green Vegetables & - & + & 40.73 \\ 
   & veg\_2 & Deep-yellow Vegetables & - & + & 14.11 \\ 
   & veg\_3 & Tomato & Low & High & 12.50 \\ 
   & veg\_4 & White Potatoes & - & + & 50.81 \\ 
   & veg\_5 & Fried Potatoes & - & + & 12.10 \\ 
   & veg\_6 & Other Starchy Vegetables & - & + & 9.48 \\ 
   & veg\_7 & Other Vegetables & Low & High & 43.35 \\ 
   & veg\_8 & Fried Vegetables & - & + & 44.15 \\ 
   & veg\_9 & Vegetable Juice & - & + & 34.48 \\ 
   & veg\_10 & Pickled Foods & - & + & 7.06 \\ 
  \hline
\end{longtable}
\endgroup

\begin{sidewaysfigure}
    \centering
    \includegraphics[width=\textwidth]{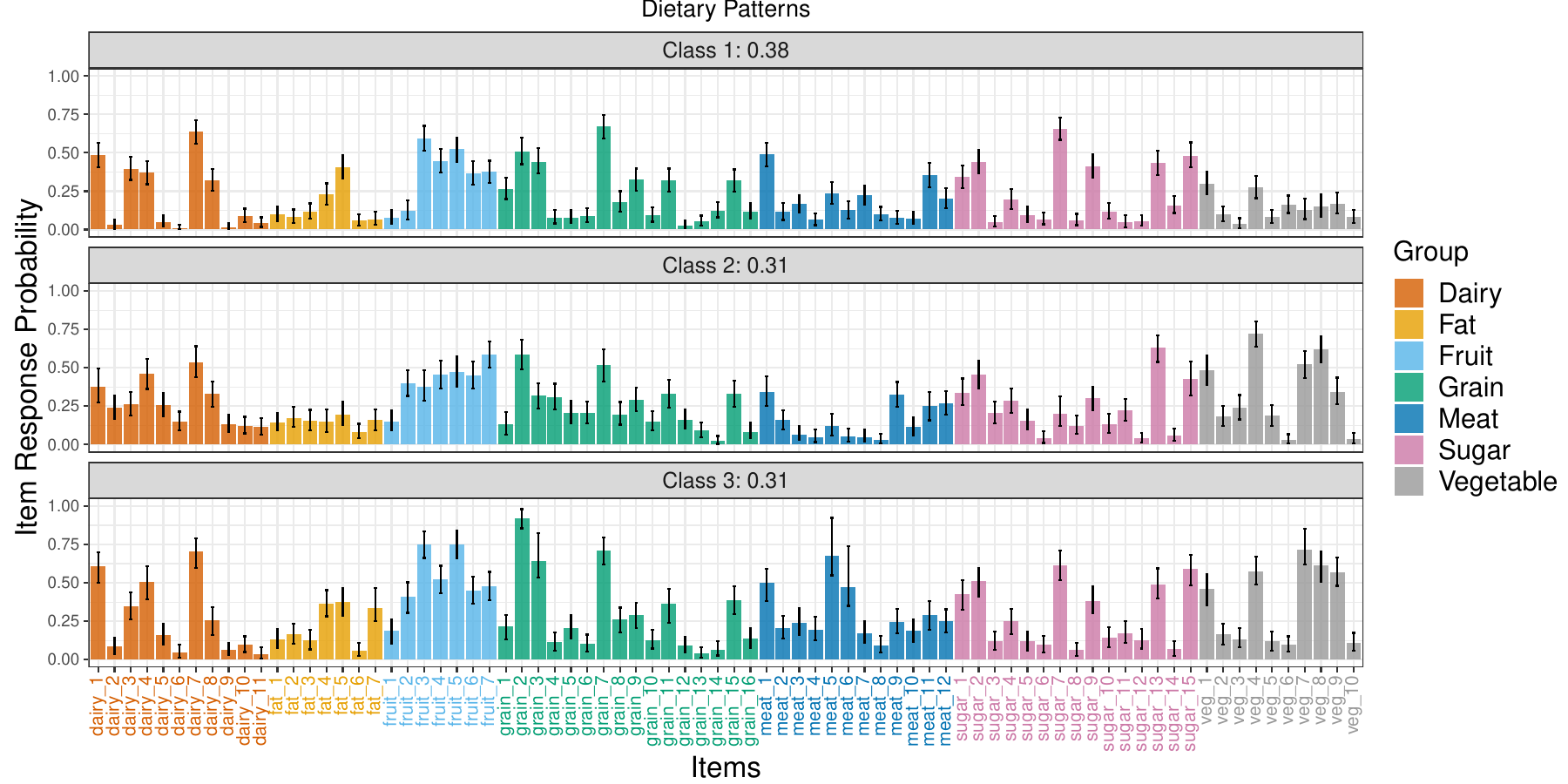}
    \caption{Results of applying Bayesian LCM to the HCHS/SOL dataset $K = 3$ latent classes.Error bars show the 95\% credible intervals from the posterior distribution. 
    }
    \label{fig:data_result:bayeslcm}
\end{sidewaysfigure}

\section{Potential Extension to Multiple Groups} \label{sec:multigroup}
In nutrition research, allowing each food item to belong to multiple food groups can be realistic in some contexts, especially for items that have diverse nutritional profiles or uses. For example, tomatoes could belong to both ``fruits'' and ``vegetables'', and avocados could belong to both ``fruits'' and ``fats''. In this case, we may aggregate the variation in the exposure probabilities of these food items across multiple groups. Specifically, suppose that we have a binary gropuing matrix $\bL = (l_{j,g})$ of size $J \times G$, where $l_{j,g}$ is 1 if item $j$ belongs to group $g$, $j = 1, \ldots, J, g = 1, \dots, G$. Note that $\bL$ is fixed according to nutrition knowledge.
Instead of restricting all items in a group to share the same variance parameter as in equation \eqref{eq:model:conddist} of the main paper, we aggregate the variances across group memberships of an item as
\begin{equation*} 
   \bm \eta_{u,j} \mid \bL, \bm \eta_{pa(u),j}, t_u, t_{pa(u)} \sim \cN \left( \bm \eta_{u,j}; \bm \eta_{pa(u),j}, \tau_j^2 (t_u - t_{pa(u)}) \right), u \in \cV \setminus \{u_0\},
\end{equation*}
where 
\begin{equation} \label{eq:multigroup:variance}
   \tau_j^2 = \sum_{g=1}^{G} l_{j,g} \sigma_g^2,
\end{equation}
with priors $\sigma_g^2 \sim$ Inv-Gamma$(a_{\sigma}, b_{\sigma}), g \in [G]$. Here we assume that there is a variance parameter $\sigma_g^2$ that controls the separability of the food group among the dietary patterns. For an individual item $j$, equation \ref{eq:multigroup:variance} combines the variability across those groups to which item $j$ belongs, into the item-specific variance $\tau^2_j$.

This parameterization accounts for the multidimensional nature of certain food items. For example, if the item $j =$ ``tomato'' belong to both ``fruits'' and ``vegetables'' groups, then whether the exposure probabilities to tomatoes are different in the $K$ dietary patterns depends on the variability of the fruit group and the vegetable group together.

\section{Discussion about Measurement Error in Dietary Surveys} \label{sec:measurementerror}
\change{
While 24-hour dietary recalls are widely used and generally considered reliable for capturing short-term intake, they are are prone to measurement errors that are common to recall data. For example, memory biases may lead to reporting food items that are actually not consumed. Portion size estimation errors may arise due to inaccurate quantificatioin of the amount of foods consumed. Misunderstanding about food composition may result in reporting incorrect food items. 

A variety of ways have been proposed to account for measurement error in model-based clustering problems. However, most of these method are designed for Gaussian mixtures using continuous observed variables; see \cite{kumar2007clustering,zhang2020model,sarkar2020gaussian,pankowska2020effect}.
To our knowledge, limited amount of work exists to deal with measurement error in analysis of multivariate discrete observations; see \cite{kreuter2008good} for a review. 
In particular, \cite{biemer2002measurement} developed an LCM with additional grouping variables (discrete covariates such as age or gender groups) to correct inconsistent reponses across repeated surveys that measure the same characteristics.
\cite{hui1980estimating} introduced separate parameters for false negative and false positive rates for each measured item. 
Latent attribute models, one type of restricted latent class models, also accounts for false positive and false negative response rates with additional model parameters \citep{gu2023joint}.
When the latent class assignment indicator derived from survey responses is used as a predictor, \cite{elliott2020methods} suggested that the uncertainty in the estimated class indicator can be viewed as absorbing measurement errors.

In dietary pattern analysis specifically, several strategies can be employed. 
First, combining different dietary assessment tools objective biomarkers can help cross-validate data and reduce reliance on a single method prone to specific errors.  For example,\cite{kipnis2003structure} discusses about correcting reporting biases in FFQ using 24-hour dietary recalls and biomarker-calibrated nutrient intake. 
Second, repeated recalls can be used to adjust for measurement errors. The NCI method pools multiple 24-hour recalls \citep{tooze2002analysis,tooze2006new} and estimates the usual instake distributions for episodically-consumed foods. The method models the probability to consume a food on a particular day and the amount eaten given that the food was consumed on a day. The two parts are linked by correlated random effects.

We may borrow similar ideas to adjust for measurement errors in our motivating recall data. However, due to limited data access to the HCHS/SOL study, we leave this direction as future work.
}



\end{document}